\documentclass[sigconf]{acmart}
\usepackage{amsmath, amsfonts, amsthm} 
 
\usepackage{amssymb}
\usepackage{mathrsfs}
\usepackage{dsfont}
\usepackage{algorithm}
\usepackage{algorithmic}
\usepackage{array}
\usepackage{cutwin}
\newtheorem{thm}{Theorem}
\newtheorem{cor}{Corollary}
\newtheorem{lem}{Lemma}
\newtheorem{defi}{Definition}

\usepackage{graphicx}
\usepackage{subfigure}
\usepackage{wrapfig}
\usepackage{cases}
\usepackage{xcolor}
\usepackage{enumerate}
\usepackage{multirow}
\usepackage{multicol}
\usepackage{setspace}
\usepackage{enumitem}
\usepackage{setspace}
\usepackage{soul}
\definecolor{aa}{RGB}{0,0,212}
\usepackage{hyperref}
\hypersetup{colorlinks=true, linkcolor=aa, anchorcolor=blue, citecolor=red, breaklinks=true}
\usepackage{courier}
\allowdisplaybreaks[4]
\usepackage{booktabs}
\usepackage{caption}

\setcopyright{none}
\renewcommand\footnotetextcopyrightpermission[1]{}

\begin{document}

\title{Will the Winner Take All? Competing Influences in Social Networks Under Information Overload}

\author{Chen Feng}
\affiliation{
  \institution{Shanghai Jiao Tong University}
  \streetaddress{School of Electronic Information and Electrical Engineering}
}
\email{ fengchen@sjtu.edu.cn}

\author{Jiahui Sun}
\affiliation{
  \institution{Shanghai Jiao Tong University}
  \streetaddress{School of Electronic Information and Electrical Engineering}
}
\email{ jhsun1997@sjtu.edu.cn}


\author{Luoyi Fu}
\affiliation{
  \institution{Shanghai Jiao Tong University}
  \streetaddress{School of Electronic Information and Electrical Engineering}
}
\email{yiluofu@sjtu.edu.cn}
\renewcommand{\shortauthors}{Feng and Fu, et al.}

\begin{abstract}
    Influence competition finds its significance in many applications, such as marketing, politics and public events like COVID-19. Existing work tends to believe that the stronger influence will always win and dominate nearly the whole network, i.e., ``winner takes all''. However, this finding somewhat contradicts with our common sense that many competing products are actually coexistent, e.g., Android vs. iOS. This contradiction naturally raises the question: \textit{will the winner take all?}
    
    To answer this question, we make a comprehensive study into influence competition by identifying two factors frequently overlooked by prior art: (1) the incomplete observation of real diffusion networks; (2) the existence of information overload and its impact on user behaviors. To this end, we attempt to recover possible diffusion links based on user similarities, which are extracted by embedding users into a latent space. Following this, we further derive the condition under which users will be overloaded, and formulate the competing processes where users' behaviors differ before and after information overload. By establishing the explicit expressions of competing dynamics, we disclose that information overload acts as the critical ``boundary line'', before which the ``winner takes all'' phenomenon will definitively occur, whereas after information overload the share of influences gradually stabilizes and is jointly affected by their initial spreading conditions, influence powers and the advent of overload. Numerous experiments are conducted to validate our theoretical results where favorable agreement is found. Our work sheds light on the intrinsic driving forces behind real-world dynamics, thus providing useful insights into effective information engineering.
\end{abstract}


\maketitle

\section{Introduction}

    Since its outbreak in December 2019, the new coronavirus disease (COVID-19) has been rapidly spreading through 216 countries and areas, causing already 67,210,778 cases and 1,540,777 deaths at the time of writing \cite{COVID19_WHO}. People are thrown into a panic for especially effective therapies, providing a hotbed for conflicting ideas, reports and information etc.. A notable example may be the controversy over the effect of sesame oil. At the beginning of 2020, a rumor that applying sesame oil to the body would block the entry of the new coronavirus is widely spreading among people. Although the World Health Organization quickly refuted such misinformation, an increased sale of sesame oil is still reported \cite{Sesame_oil}. Competitions between conflicting ideas (such as whether to prevent COVID-19 by taking a hot bath, drinking alcohol and eating garlic \cite{WHO_mythbusters}) continue to happen through the year, significantly affecting people's life.

\begin{figure}[t]
    \centering
      \subfigure[Facebook vs. Myspace]
        {
        \begin{minipage}[h]{0.232\textwidth}
        \vspace{-.7mm}
            \includegraphics[width=1\textwidth]{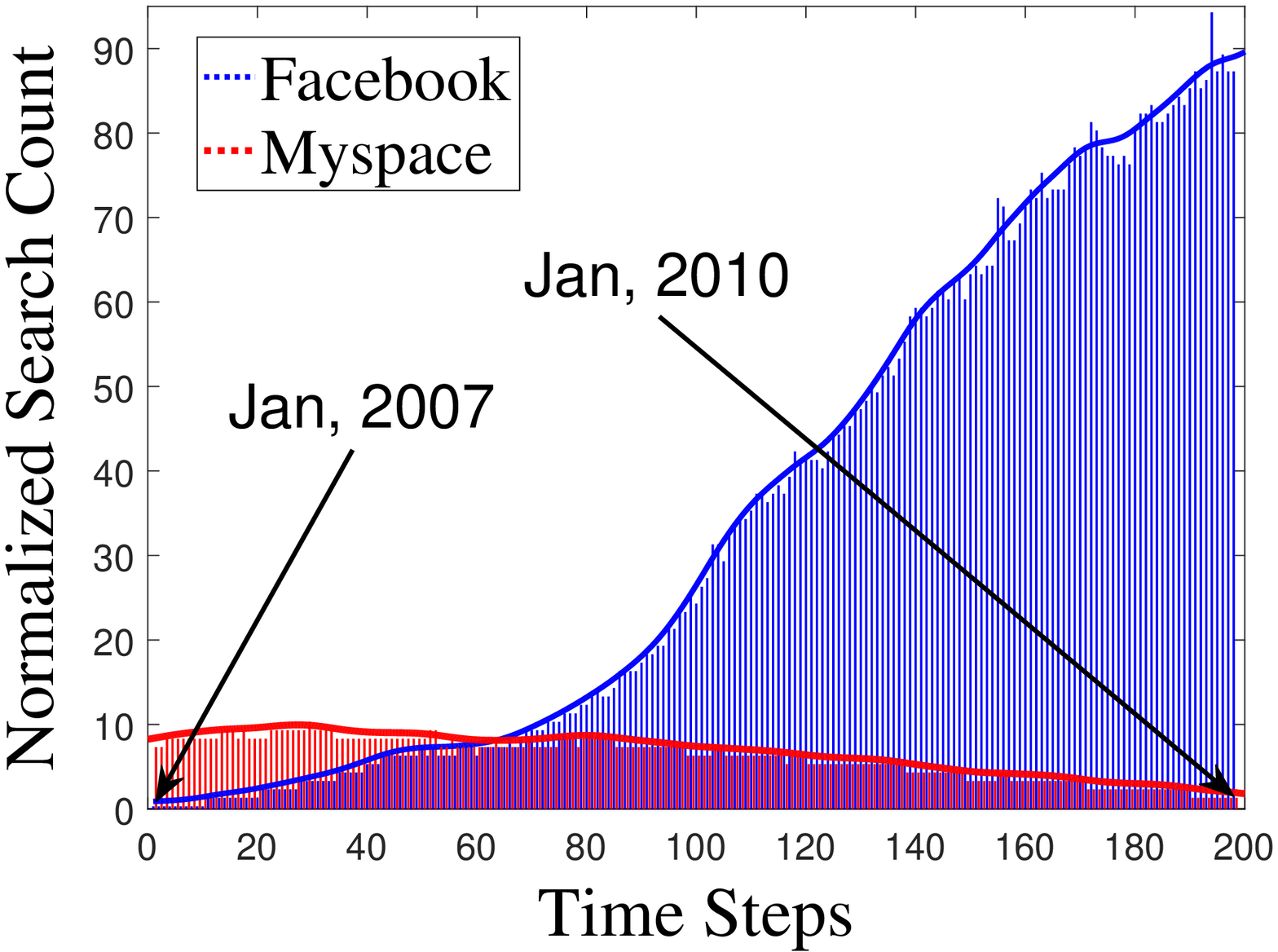}\label{fig_facebook_myspace}
            \vspace{-3.2mm}
          \end{minipage}}
      \hspace{-2mm}  
      \subfigure[Android vs. iOS]
      {
      \begin{minipage}[h]{0.212\textwidth}
        \vspace{-.5mm}
           \includegraphics[width=1\textwidth]{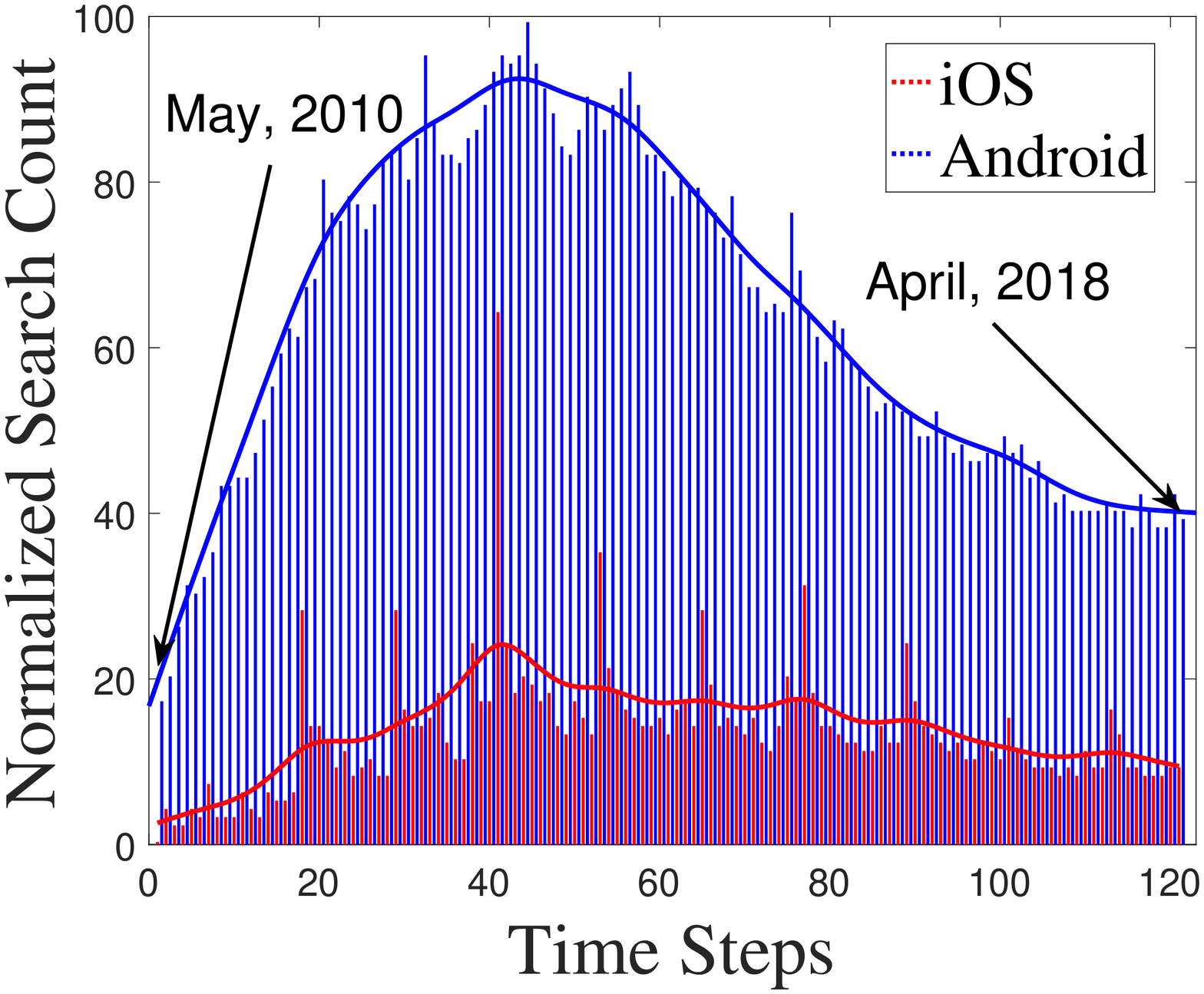}\label{fig_android_ios}
           \vspace{-3.2mm}
         \end{minipage}}
      \vspace{-4.5mm}
      \caption{Search interests of competing products over time\protect\footnotemark[2]. (a) In the early years, Myspace is shown to have an edge over Facebook. However, as time goes on, Facebook  gradually manifests its advantage and finally dominates nearly the whole market. (b) Across the ten years' competition between Android and iOS, although Android is shown to be more popular, it does not dominate the whole market, refuting the ``winner takes all'' phenomenon.}\label{fig_competing_products_intro}
      \vspace{-5mm}
    \end{figure}

    Should we have deeper insights into the competing dynamics of conflicting information, the incidents could have been avoided by accurately engineering the propagation process. Besides pandemic control, similar demands are raised in many other areas such as marketing, politics and social ideology. To fill this gap, a body of work has been dedicated to studying the competition between conflicting information, products, opinions etc.\protect\footnotemark[1]. The work of Prakash et al. \cite{WWW_winner} marks a milestone in investigating the problem, by revealing the well-known ``winner takes all'' phenomenon in the marketing theory, that the stronger influence will become the winner and finally wipe out the weaker, as shown in Fig. \ref{fig_facebook_myspace}. Numerous subsequent researches reach the same conclusion that only one influence will survive in the final equilibrium state \cite{Basar_Tamer_bi_virus} \cite{bi_virus_generic_rate} \cite{to_live_or_to_die} \cite{JSAC_competing_memes}. However, this finding somewhat contradicts with our daily experience, since many competing products are observed to be coexistent, such as McDonald's vs. KFC, Boeing vs. Airbus and Android vs. iOS, as illustrated in Fig. \ref{fig_android_ios}. There is also little literature showing that competing influences will always co-exist. Especially, Koprulu et al. find that the market share of products will gradually stabilize and is basically proportional to their persuasive powers \cite{opinion_battle}, i.e., winner will not take all. The contradictory conclusions and opposite examples naturally lead to the question: \textit{will the winner take all?}
    \footnotetext[1]{For ease of appellation, we refer to them all as ``influences''.}

    \footnotetext[2]{The data is available on Google Trends \cite{google_trends}. In each subgraph, the search count is normalized by the largest count.}

    \textbf{Competition Review.} To answer this question, we may need to review the competing process for an overall view. In essence, the competition starts with the injection of influences. Then, the influences propagate along the given network. And the users therein decide which influence to follow. As can be seen, the competition process is an interplay of three parts: the influences themselves (\textit{intrinsic factor}), the users (\textit{subjective factor}) and the underlying networks (\textit{objective factor}). It is worth noting that previous work has delicately considered the intrinsic factor via various characterizations, such as propagation rate \cite{WWW_winner}, persuasive capability \cite{opinion_battle}. However, the latter two factors are frequently overlooked in existing literature. To elaborate, (1) \textit{\underline{subjective factor:}} most work ideally assumes a fixed user behavior throughout the whole competing process. However, in reality, as time goes on, more users will become influenced and turn a spreader. Thus, a late decision maker would receive more information which possibly exceeds users' attention limit and leads to information overload \cite{scientific_report_limited_attention}. Further studies reveal that overloaded users exhibit a different behavior pattern from before -- preferring priority strategies to cope with the overload \cite{quantifying_information_overload}. (2) \textit{\underline{Objective factor:}} it is often implicitly assumed in previous work that the influence propagates through a given network. However, due to users' privacy concern, the large scale of social networks etc., the observed network is often only a subgraph of real networks. Many legitimate links are actually not detected.

    \textbf{The New Formulation.} In this work, we take all the three factors into account. Specifically, we consider two competing influences $I_1$ and $I_2$ propagating through an observed social network $G(V,E)$. The influences are characterized by their influence powers $a$ and $b$ respectively. Each user is featured by their capacity in handling information, i.e., $\delta_c$. Initially, users have enough attention to discriminate the influences and tend to follow influences by the received powers. Beyond threshold $\delta_c$, users are generally overloaded and four typical information processing strategies are considered for users to cope with the overload, with each defined according to some priority, e.g., the arrival time.
    


    To circumvent the incompleteness of the observed network, we look into the generation of social relationships. It is found in social science that network connections generally stem from user similarities \cite{birds_2} and users tend to interact with similar ones \cite{birds_of_a_feather}, known as the ``homophily principle''. Thus, we are inspired to recover possible diffusion links based on user similarities,  before running the competition. In the attempt to extract user similarities, we find that this idea coincides with an emerging technique in link prediction called \textit{Network Embedding} \cite{node2vec} \cite{Embedding_link_prediction_1} \cite{Embedding_link_prediction_2}, where users are mapped into a latent space with user similarities encoded as their distances. 

    \textbf{Technical Challenges.} Proceeding to reveal the competing dynamics, we are confronted with three major challenges. 
    
    \textbf{$\bullet$} User embeddings are typically concerned with practical applications but hardly analytically captured, leaving properties of the latent space unknown (esp. distance distribution) which however are an essential prerequisite for diffusion analyses.
    
    \textbf{$\bullet$} Four information processing strategies are defined in a broad way for generality, while imposing substantial difficulties in mathematical characterization of user behaviors and accordingly the analysis of competing processes.
     
    \textbf{$\bullet$} Like existing work, in revealing the competing dynamics, we are faced with the long-standing difficulty of deriving explicit expressions for each influence w.r.t. each time instant.

    \textbf{Our Solution.} To meet the above challenges, we first estimate users' distribution in the latent space and further disclose the distribution of user distances. On this basis, the time of overload is derived. Then, the diffusion processes are formalized as differential equations, based on the characterization of the four information strategies. Finally, we solve the equations and reveal the underlying competing dynamics.


    Numerous contributions are made when fulfilling this work, among which the highlights are summarized as follows.

    \textbf{$\bullet$} This work acts as a comprehensive study of influence competition, identifying not only intrinsic factor but also subjective factor and objective factor in the competing process.

    \textbf{$\bullet$} To our knowledge, our work is the first attempt to mathematically delineate the properties of latent space. Moreover, in the case of overload, users' behaviors under four strategies are analytically captured in a unified form. 

    \textbf{$\bullet$} We manage to derive the competing dynamics in explicit expressions, showing that the winner will not always ``take all''. To explain, before overload, the stronger influence will be the winner and dominate the network, exhibiting the ``winner takes all'' phenomenon. While under information overload, the result is jointly determined by the initial state, influence power and the time of overload. And the weaker influence could even become the winner and maintain a modest share of market, given enough initial followers and proper overload timing.
    

    \textbf{$\bullet$}  Numerous experiments are carried out on real-world networks and the results are found to be in favorable agreement with our theoretical findings.

    
    \textbf{Roadmap.} We formulate the problem in Sec. \ref{sec_formulation} and study the properties of the latent space in Sec. \ref{sec_property_of_latent_space}. Then, we reveal the competing dynamics before and after overload in Sec. \ref{sec_before_overload} and \ref{sec_after_overload}. The theoretical findings are further validated in Sec. \ref{sec_experiment}. Finally, we conclude the paper in Sec. \ref{sec_conclusion}.
    
    In Sec. \ref{sec_discussion}, we answer the question raised in the title and discuss the implications of our findings in practical events like COVID-19.

\section{Problem Formulation}\label{sec_formulation}

   The observed social network is denoted as $G(V,E)$, where $V$ is the set of users ($|V|=n$) and $E$ records user connections. Users' degree is assumed to follow the power law distribution with parameter $\lambda>1$, which is widely observed in the real world \cite{nature_power_law} \cite{power_law_2}. Then, the probability of a user having degree $k$ is $C/k^\lambda$, where $C$ is the normalization factor.
    
   Two competing influences $I_1$ and $I_2$ are spreading among users, with influence powers $a$ and $b$ respectively. The competing process starts at time $t_0$, with initial states $X_1(t_0)$ and $X_2(t_0)$,\protect\footnotemark[3] where  $X_1(t)$ (resp. $X_2(t)$) denotes the number of users affected by $I_1$ (resp. $I_2$) at time $t$ and $X_1(t)+X_2(t)=X(t)$. Then, the share of each influence could be expressed as $X_1(t)/X(t)$ and $X_2(t)/X(t)$.
  \footnotetext[3]{Without loss of generality (w.l.o.g.), we assume $t_0$ is equal to $X(t_0)$ in value. Then, according to the diffusion mechanism defined later, at any time step, $t$ is exactly the number of users infected, i.e., $X_1(t)+X_2(t)=t$.\label{footnote_initial_users}}

\subsection{Network Embedding}

  In practice, the subsequent diffusion process is not likely to expand along the observed network exactly, since many legitimate links are not observed due to users' privacy concern, the large scale of real networks etc.. In this regard, we intend to recover possible diffusion links based on user similarities, since social connections often originate from user similarities \cite{birds_2} and users tend to interact with similar ones \cite{birds_of_a_feather}.


  To this end, we adopt the well-known ``node2vec'' technique \cite{node2vec} to abstract user similarities due to its high accuracy in link prediction (typically over 90\%). Specifically, users are mapped into a $d$ dimensional latent space, where each user $i$ is represented by a vector $\mathbf{v}_i\in \mathbb{R}^d$ and the similarity between users $i,j$ is encoded by their distance $l(i,j)$.
  
  The framework of ``node2vec'' mainly consists of two parts: neighborhood sampling and embedding optimization.
  
  \textbf{Neighborhood Sampling.} For each user $i\in V$, we stimulate a biased random walk parameterized by $p$ and $q$, as defined in Definition \ref{Def_node2vec_walk}, to obtain her wide-sense neighborhood $N_S(i)$.
  
    \begin{defi}\label{Def_node2vec_walk}
    \setlength{\abovedisplayskip}{3pt}
    \textbf{(Sampling Strategy)} Consider a random walk which just traversed edge $(u,v)$ and reaches node $v$. Then, neighbors of $v$ are assigned weight 1 (esp. $u$ itself is assigned $1/p$), and others are of weight $1/q$. The walk next reaches the neighbors of $v$ with probability proportional to their weights.
    \setlength{\abovedisplayskip}{3pt}
  \end{defi}
  
  \textbf{Embedding Optimization.} 
  We next seek to maximize the log-probability of observing each neighbor pair in the latent space
  \begin{equation}
  \setlength{\abovedisplayskip}{3pt}
    \mathcal{O}=\sum_{i\in V}\sum_{j\in N_S(i)}\log P(i,j).
  \setlength{\abovedisplayskip}{6pt}
  \end{equation}
   Specifically $P(i,j)$ is a softmax function w.r.t. user distance
  \begin{equation}\label{eq_conditional_prob_of_each_pair}
  \setlength{\abovedisplayskip}{3pt}
    P(i,j)=\frac{ \exp(-l(i,j)) }  { \sum_{w\in V}\exp(-l(i,w)) },
  \setlength{\abovedisplayskip}{6pt}
  \end{equation}
  where $l(i,j)\!=\!\|\mathbf{v}_i-\mathbf{v}_j\|_2^2$, i.e., the squared Euclidean distance, in our work.
  Accordingly, we obtain the objective function
  \begin{equation}\label{eq_objective_likelihood}
  \setlength{\abovedisplayskip}{3pt}
    \mathcal{O}=-\!\sum_{i\in V}\big[\!\sum_{j\in N_S(i)}\!\!\!\!l(i,j)+|N_S(i)|\log\!\! \sum_{w\in V}\!\!\exp(-l(i,w))\big].
  \setlength{\abovedisplayskip}{6pt}
  \end{equation}
  Finally, the objective $\mathcal{O}$ could be optimized by applying the stochastic gradient ascent algorithm.
  
  For an analytical characterization, user embeddings are captured by their distribution in the latent space $h(\mathbf{v}_{\cdot},\theta)$, where the parameter $\theta$ is estimated by estimation methods like maximum likelihood.

\subsection{Competing Mechanism}

  Based on the given network and user embeddings, we could derive the recovered diffusion network $G(V,E^\prime)$, where the edge set $E^\prime$ is the union of the original social relations $E$ and the latent links recovered from similarities. Specifically, we assume a latent link would exist between two users if their distance in the latent space is smaller than $r\in \mathds{R^+}$, since users are not likely to interact with ones not similar enough. Correspondingly, users' neighborhood consists of both neighbors in $G(V,E)$ and users within $r$.
   
  On this basis, the competing process unfolds in discrete steps as follows. In each step, a random user $i$ (called \textit{riser}) will arise to make a decision on which influence to follow, based on the influences she receives from her neighbors (if none of her neighbors is influenced, it is natural to assume she will not arise). 

\begin{figure}[h]
      \centering
        \vspace{-0.2cm}
      \includegraphics[width=0.45\textwidth]{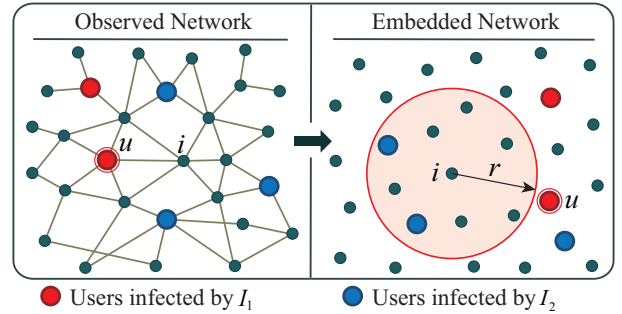}
        \vspace{-0.3cm}
      \caption{An illustration of the influence propagation model.}\label{Fig_inf_propagation_model}
        \vspace{-2mm}
    \end{figure}

  In the early stage, since the influenced users are relatively rare, the riser generally will not receive too much information and become overloaded. Therefore, she would be able to discriminate the power of influences and will follow influences with probability proportional to the cumulative influence power received (just like \cite{JSAC_competing_memes}), as illustrated in Fig. \ref{Fig_inf_propagation_model}. To elaborate, at time $t$, user $i$ arises to make a decision and finds that three of her neighbors are influenced with one original neighbor (user $u$) in red, two latent neighbors in blue. Then, user $i$ will become red with probability $\frac{a\times 1}{a\times 1+b\times 2}=\frac{a}{a+2b}$, and blue with probability $\frac{2b}{a+2b}$. 

  As time goes on, more users will become influenced. And a new riser will receive more information when making a decision. The abundance of information will finally exceed user's capacity to consume (denoted as $\delta_{c}$), since user's attention is limited by \cite{scientific_report_limited_attention}. Being overloaded with information, users may not have enough attention to carefully identify the influences and tend to adopt priority strategies to cope with the overload, according to the studies in \cite{jacob_overload} \cite{quantifying_information_overload}. 
  
  To capture this phenomenon, we assume an overloaded user will discriminate the influences as before with probability $p(t)=e^{-\mu (t-t_{c})}$ ($t>t_c$), which decreases exponentially over time with a factor of $\mu$, where $t_{c}$ is the time when users are generally overloaded, i.e., receiving too much information that exceeds users' capacity. On the other hand, with probability $1-p(t)$, users would turn to priority strategies to help process information. Especially, four typical strategies are considered in Definition \ref{Def_overloaded_behaviors}.
  \begin{defi}\label{Def_overloaded_behaviors}
    \textbf{(Overloaded Behaviors)} Under information overload, we consider four generic information processing strategies for users to cope with the overload.
    \begin{itemize}
      \item[(i)] The riser adopts the first influence that reaches her, given arbitrary distribution of information propagation time \cite{to_live_or_to_die}.
      \item[(ii)] Contrary to (ii), the riser adopts the latest (i.e., the newest) influence, regardless of the distribution of propagation time.
      \item[(iii)] The riser follows the influence that her most similar neighbor chooses \cite{birds_of_a_feather}, under any user distributions in the latent space.
      \item[(iv)] The riser follows the influence her most popular neighbor (i.e., having the largest degree) adopts, under any degree distribution of the network.
    \end{itemize}
  \end{defi}

\subsection{Problem Statement}
    We are now ready to state the problem formally. Given an observed social network $G(V,E)$, user distribution $h(\mathbf{v}_\cdot;\theta)$ in $\mathbb{R}^d$ and initial states $X_1(t_0)$ and $X_2(t_0)$, influences $I_1$ and $I_2$ with powers $a$ and $b$ are propagating among users according to the competing mechanism characterized by parameters $r$, $\delta_{c}$ and $\mu$, where in practice the parameters could be derived by experimental estimation. 

    The goal is to derive the number of users infected by $I_1$ and $I_2$ with regard to each time step (i.e., $X_1(t)$ and $X_2(t)$) and thus answer the question ``\textit{will the winner take all?}''.

\section{Properties of the Latent Space}\label{sec_property_of_latent_space}

  In this section, we investigate the properties of the latent space. To obtain explicit results, we apply the $d$ dimensional Gaussian distribution $\mathcal{N}(\mathbf{u},\sigma^2 \mathbf{I})$ to depict depict user embeddings like many previous studies \cite{Embedding_link_prediction_1} \cite{Embedding_link_prediction_2}, where $\mathbf{u}$ is a $d\times 1$ mean vector, $\sigma^2$ is the variance and $\mathbf{I}$ is the $d\times d$ identity matrix. The parameters $\mathbf{u}$ and $\sigma^2$ could be estimated by the maximum likelihood method.

  \begin{defi}\label{Def_gamma_dist}
    \textbf{(Gamma Distribution)} A random variable $Z$ follows the Gamma distribution $\Gamma(\alpha,\beta)$, if its probability density function $f(z;\alpha,\beta)$ satisfies
      \begin{equation}
      \setlength{\abovedisplayskip}{3pt}
        f(z;\alpha,\beta)=
        \left\{
          \begin{aligned}
             &\frac{  \beta^\alpha  } {  \Gamma(\alpha)  } z^{\alpha-1}  e^{-\beta z}  ,&&z\geq 0,\\
             &0, &&z<0.\\
          \end{aligned}
        \right.
        \setlength{\abovedisplayskip}{3pt}
      \end{equation}
      where $\Gamma(\alpha)=\int_0^\infty z^{\alpha-1}e^{-z}dz$ is the gamma function.
  \end{defi}

  Let random variable (r.v.) $L$ denote the squared Euclidean distance between a random pair of users. We next examine the distribution of $L$ and then derive the probability that the distance between two users is within $r$ by Lemmas \ref{Lemma_distance_dist} and \ref{Lemma_within_R} respectively, where the proofs are deferred to the technical report \cite{TRRRRR} for readability.

    \begin{lem}\label{Lemma_distance_dist}
      Given the distribution of user positions $\mathcal{N}(\mathbf{u},\sigma^2 \mathbf{I})$, the squared Euclidean distance between users $L$ follows the Gamma distribution $\Gamma(\frac{d}{2},\frac{1}{4\sigma^2})$.
    \end{lem}

  \begin{lem}\label{Lemma_within_R}
    In the latent space with user's position distribution $\mathcal{N}(\mathbf{u},\sigma^2 \mathbf{I})$, the probability that the distance between two random users $i,j$ is smaller than $r$ (i.e., $l(i,j)\leq r$) is
    \begin{equation*}
      p(r,\sigma^2)=\frac{1}{\Gamma(\frac{d}{2})}\int_0^{\frac{r}{4\sigma^2}}  z^{\frac{d}{2}-1}e^{-z} dz,
    \end{equation*}
    which increases with influence range $r$ and decreases with position variances $\sigma^2$.
  \end{lem}

  Note that $p(r,\sigma^2)$ is an incomplete gamma function, there is no closed form expression to calculate it. However, we could approximate it with the following formula according to \cite{approx_incomplete_gamma_func}
    \begin{equation*}
      \Gamma(\frac{d}{2})-\gamma(\frac{d}{2},\frac{r}{4\sigma^2}),
    \end{equation*}
  where function $\gamma(a+1,x)$ is defined as $\frac{e^{-x}x^{a+1}}{x-a}[1-\frac{a}{(x-a)^2}+\frac{2a}{(x-a)^3}+O(\frac{a^2}{(x-a)^4})]$.

  With the above results, we are ready to derive the time when new risers will generally feel information overload, i.e., the number of infected neighbors exceeds the threshold $\delta_{c}$.

  \begin{thm}\label{Thm_overload_time}
    With probability $1-o(1)$, a new riser will feel information overload if she arises after time $t_{c}=\frac{\delta_{c}}{(1+o(1))p(r,\sigma^2)}$.
  \end{thm}
  \begin{proof}
    Consider a random user $i$ with degree $d_i$ in $G(V,E)$. We first examine the number of her actual neighbors, which consists of two parts: the neighbors in the original network and the latent neighbors due to similarity. For the first case, We would like to show that $d_i=o(n)$ with high probability. Recall that $d_i$ follows the power law distribution parameterized by $\lambda>1$. When $\lambda\in (1,2)$, the expectation of user degree is $\mathds{E}(d_i)=\int_1^n x\frac{C}{x^\lambda} dx=\Theta(n^{2-\lambda})$. By Chernoff bound, we have
    \begin{equation}\label{eq_chernoff}
        P(|X-\mathds{E}(X)|\leq \epsilon \mathds{E}(X))\geq 1-2e^{-\mathds{E}(X)\epsilon^2/3}.
    \end{equation}
    Let $\epsilon$ be $\sqrt{\frac{3\log n}{n^{2-\lambda+\epsilon^\prime}}}$, we have that $d_i$ concentrates around its expectation, which is inferior to $n$ since $\lambda>1$, with probability $1-o(1)$. Similarly, when $\lambda \geq2$, $\mathds{E}(d_i)=O(\log n)$. By setting $\epsilon=\log n$, we see that $d_i<(\log n)^2=o(n)$ with probability $1-o(1)$.
    
    For the second case, the number of her latent neighbors is a random variable (denote it as $X$) which follows a binomial distribution $B(n-d_i,p(r,\sigma^2))$. Applying the Chernoff bound in Eq. (\ref{eq_chernoff}) by setting $\epsilon=\sqrt{\frac{3\log n}{(n-d_i)p(r,\sigma)}}$, we see that $X=(1+o(1))\mathds{E}(X)$ with probability at least $1-\frac{1}{(1+o(1))n}$ which approaches 1 as $n\rightarrow \infty$. Thus, we could approximate $i$'s latent neighborhood with its expectation $(n-d_i)p(r,\sigma^2)$.  

    To summarize, the size of user's actual neighborhood is $d_i+(n-d_i)p(r,\sigma^2)=(1+o(1))np(r,\sigma^2)$. Then, at time $t$, the expected number of infected neighbors is $(1+o(1))np(r,\sigma^2)\frac{t}{n}$. Setting it to be $\delta_c$, we derive the time when users will be overloaded $t_c=\frac{\delta_{c}}{(1+o(1))p(r,\sigma^2)}$.

  \end{proof}
  \textbf{Remark.} The theoretical results agree well with our common intuitions. To explain, Theorem \ref{Thm_overload_time} implies that the time point of overload is directly proportional to users' ability of handling information (i.e., $\delta_c$), and inversely proportional to the probability of having a neighbor, i.e., $p(r,\sigma^2)$. Taking a closer look into $p(r,\sigma^2)$, by Lemma \ref{Lemma_within_R} we find that if users are more sociable, i.e., with a larger $r$, they tend to have more neighbors and are thus more likely to be overloaded. On the other hand, in a sparsely connected network, implying a larger $\sigma^2$, users often have fewer friends, resulting in a smaller probability of being overloaded. 
  
  On this basis, we could infer that in $G(V,E^\prime)$ users are more likely to be overloaded and have a smaller $t_c$ than in $G(V,E)$, since the recovered network has more edges (thus smaller $\sigma^2$) than the observed one (i.e., $|E^\prime|\geq |E|$).

  \vspace{1mm}
  Since user's behaviors before and after overload are different, in subsequent sections, we will analyze the two scenarios respectively.

\section{Influence Propagation Before Overload}\label{sec_before_overload}

  In this section, we consider the influence propagation before time $t_{c}$, i.e., users are not overloaded with information.

  Recall that the number of users influenced by $I_1$ and $I_2$ at time $t$ is $X_1(t)$ and $X_2(t)$ respectively. At the next time step $t+1$, suppose user $i$ arises to make a decision. In expectation, she will receive $[d_i+(n-d_i)p(r,\sigma)]X_1(t)/n$ pieces of information about $I_1$ and $[d_i+(n-d_i)p(r,\sigma)]X_2(t)/n$ pieces of information about $I_2$. Then, according to the competing mechanism, we have two possible outcomes of $i$'s decision:
  \begin{itemize}[noitemsep,topsep=0pt]
  \begin{spacing}{1.2}
    \item user $i$ gets influenced by $I_1$ with probability $\frac{aX_1(t)}{aX_1(t)+bX_2(t)}$;
    \item user $i$ gets influenced by $I_2$ with probability $\frac{bX_2(t)}{aX_1(t)+bX_2(t)}$.
  \end{spacing}
  \end{itemize}
  \vspace{-4mm}

  Given current states $X_1(t)$ and $X_2(t)$, the mean state of the system at time $t+1$ could be characterized by the following difference equation set
  \begin{equation}\label{eq_difference_eq_before_overload}
    \setlength{\abovedisplayskip}{3pt}
      \left\{
          \begin{aligned}
             &\mathds{E}[X_1(t+1)-X_1(t)|X_1(t)]=\frac{aX_1(t)}{aX_1(t)+bX_2(t)},\\
             &\mathds{E}[X_2(t+1)-X_2(t)|X_2(t)]=\frac{bX_2(t)}{aX_1(t)+bX_2(t)}.\\
          \end{aligned}
        \right.
    \setlength{\abovedisplayskip}{3pt}
  \end{equation}

  To solve the equations, we present an equivalent view to the propagation process in a continuous way and convert the difference equations into differential equations. Specifically, we suppose the decision process expands as follows
  \begin{equation*}
  \setlength{\abovedisplayskip}{3pt}
    \begin{aligned}
      &P(\text{A random user makes decision during }(t,t+\tau])=\tau,\\
      &P(\text{No user makes decision during }(t,t+\tau])=1-\tau,
    \end{aligned}
    \setlength{\abovedisplayskip}{3pt}
  \end{equation*}
  where $\tau\in (0,1]$. Then, the number of infected users at time $t$ is the same as the original discrete model. For this continuous model, the first equation in Eq. (\ref{eq_difference_eq_before_overload}) could be written as
  \begin{equation}\label{eq_CT_difference_sample}
  \setlength{\abovedisplayskip}{3pt}
    \mathds{E}[X_1(t+\tau)-X_1(t)|X_1(t)]=\tau\frac{aX_1(t)}{aX_1(t)+bX_2(t)}.
  \setlength{\abovedisplayskip}{6pt}
  \end{equation}

  Let $x_1(t)$ (resp. $x_2(t)$) denote the expected number of users infected by $I_1$ (resp. $I_2$) in the continuous model, i.e., $x_1(t)=E(X_1(t))$ (resp. $x_2(t)=E(X_2(t))$). Taking the expectation of Eq. (\ref{eq_CT_difference_sample}), we obtain
    \begin{equation}\label{eq_CT_expected}
    \setlength{\abovedisplayskip}{6pt}
      x_1(t+\tau)-x_1(t)=\tau\frac{ax_1(t)}{ax_1(t)+bx_2(t)}.
    \setlength{\abovedisplayskip}{3pt}
    \end{equation}
  Since Eq. (\ref{eq_CT_expected}) holds for $\tau\in (0,1]$, we have
    \begin{equation}
    \setlength{\abovedisplayskip}{3pt}
      \lim_{\tau\rightarrow 0}\frac{x_1(t+\tau)-x_1(t)}{\tau}=\frac{ax_1(t)}{ax_1(t)+bx_2(t)}.
    \setlength{\abovedisplayskip}{3pt}
    \end{equation}
  Then, we conduct similar processes to the second equation in Eq. (\ref{eq_difference_eq_before_overload}). Finally, by definition of the derivative, we obtain the differential description of the propagation process
    \begin{equation}\label{eq_differential_eq_before_overload}
      \left\{
          \begin{aligned}
             &\frac{dx_1}{dt}=\frac{ax_1(t)}{ax_1(t)+bx_2(t)},\\
             &\frac{dx_2}{dt}=\frac{bx_2(t)}{ax_1(t)+bx_2(t)}.\\
          \end{aligned}
        \right.
    \end{equation}

    On this basis, we obtain the influence dynamics before information overload by Theorem \ref{Thm_influence_before_overload}.
    \begin{thm}\label{Thm_influence_before_overload}
      At time $t$, the number of users infected by $I_1$ and $I_2$ satisfies the following equations respectively
      \begin{equation}\label{eq_thm_results_before_overload}
        \left\{
        \begin{aligned}
          &C_1x_1^{\frac{b}{a}}(t)+x_1(t)=t,\\
          &C_2x_2^{\frac{a}{b}}(t)+x_2(t)=t,\\
        \end{aligned}
        \right.
      \end{equation}
      where $C_1=[x_2(t_0)]/{x_1^{\frac{b}{a}}(t_0)}$, $C_2=[x_1(t_0)]/{x_2^{\frac{a}{b}}(t_0)}$.
    \end{thm}
    \begin{proof}
      We obtain the expressions about $x_1(t)$ and $x_2(t)$ by solving the differential equations in Eq. (\ref{eq_differential_eq_before_overload}). Note that $x_1(t)+x_2(t)=t$ at any time. For the first equation of Eq. (\ref{eq_differential_eq_before_overload}), by substituting $x_2(t)$ with $t-x_1(t)$, we have
      \begin{equation}\label{eq_change_variable_x_2}
        \frac{dx_1}{dt}=\frac{ax_1(t)}{(a-b)x_1+bt}.
      \end{equation}
      We next take the reciprocal of each side, and denote $q=\frac{t}{x_1}$. Since $q$ is also a function of $x_1$, we derive that $\frac{dt}{dx_1}=q+x_1\frac{dq}{dx_1}$. Then, Eq. (\ref{eq_change_variable_x_2}) is equivalent to
      \begin{equation*}
        x_1\frac{dq}{dx_1}=\frac{a-b}{a}+(\frac{b}{a}-1)q.
      \end{equation*}
      By separating the variables, we have
      \begin{equation*}
        \frac{a}{a-b}\frac{1}{1-q}dq=\frac{1}{x_1}dx_1.
      \end{equation*}
      Integrating both sides and taking anti-logarithm, after some simplifications, we obtain
      \begin{equation*}
        C_1x_1^{\frac{b}{a}}(t)+x_1=t.
      \end{equation*}
      where $C_1$ is a constant. Solving $C_1$ with the initial condition $x_1(t_0)$ and $x_1(t_0)+x_2(t_0)=t_0$, we have $C_1=[{x_2(t_0)}]/{x_1^{\frac{b}{a}}(t_0)}$.

      Following the similar idea, we could obtain the expression about $x_2(t)$. And this completes the proof.
    \end{proof}

    To derive more specific expressions of $x_1(t)$ and $x_2(t)$, we study two special cases in Corollary \ref{cor_specific_expression}.
    \begin{cor}\label{cor_specific_expression}
      (i) When $\frac{a}{b}=\frac{1}{2}$, at time $t$, the number of users infected by $I_1$ (resp. $I_2$) is
        \begin{equation*}
          \left\{
            \begin{aligned}
              &x_1(t)=\frac{1}{2C_1}\left(\sqrt{4C_1 t+1}-1\right),\\
              &x_2(t)=t-\frac{1}{2}\left(C_2\sqrt{4t+C_2^2}-C_2^2\right).\\
            \end{aligned}
          \right.
        \end{equation*}
        Symmetric results could be derived when $\frac{a}{b}=2$.\\
      (ii) When $\frac{a}{b}=1$, at time $t$, the number of users infected by $I_1$ and $I_2$ is
      \begin{equation*}
        x_1(t)=\frac{x_1(t_0)}{t_0}t,\;\; x_2(t)=\frac{x_2(t_0)}{t_0}t.
      \end{equation*}
        That is to say, the influence spread of $I_1$ and $I_2$ is totally decided by the initial states.
    \end{cor}

    \textbf{Remark.} By Theorem \ref{Thm_influence_before_overload}, we find that before information overload, the competing dynamics exhibit the ``winner takes all'' phenomenon. To explain, we assume that $I_1$ is stronger (i.e., $a>b$) w.l.o.g.. Then, Eq. (\ref{eq_thm_results_before_overload}) could be written as
      \begin{equation}\label{eq_variation_thm_results_before_overload}
          x_1(t)=\frac{t}{1+C_1x_1^{\frac{b}{a}-1}(t)},     \quad          x_2(t)=\Big[\frac{t}{C_2+x_2^{1-\frac{a}{b}}(t)}\Big]^{\frac{b}{a}}.
      \end{equation}

      Note that $x_1(t)$ and $x_2(t)$ both increase over time. Thus, as $t\rightarrow \infty$, $C_1x_1^{\frac{b}{a}-1}(t)$ is an infinitesimal $o(1)$ and accordingly $x_1(t)$ scales like $t$ by Eq. (\ref{eq_variation_thm_results_before_overload}). Similarly, $x_2^{1-\frac{a}{b}}(t)$ is also in $o(1)$ and $x_2(t)$ scales like $t^{\frac{b}{a}}=o(t)$, i.e., $x_2(t)=o(x_1(t))$. Thus, we see that, as time goes on, if no overload is incurred ($t_c\rightarrow \infty$), $I_1$ will win the competition and dominate nearly the whole network, i.e., the winner takes all. Especially, if $a=b$ (i.e., no power difference), $I_1$ and $I_2$ will share the network according to their ratio in the initial states.


\section{Influence Propagation After Overload}\label{sec_after_overload}

    In this section, we proceed to study the competition at time $t>t_{c}$ when users have been overloaded. To this end, we first look into users' behavior under information overload. Then we formulate and reveal the competing dynamics of $I_1$, $I_2$.

\subsection{User Behavior Under Information Overload}

    Recall that four information processing strategies are considered in the formulation of our problem in Definition \ref{Def_overloaded_behaviors}. 
    By ordinal, we first study strategies (i) and (i), where users adopt the earliest and latest influence respectively. The corresponding infection probability of a new riser is presented in Lemmata \ref{lem_adoption_prob_mode_1} and \ref{lem_adoption_prob_mode_2}, where the proof of Lemma \ref{lem_adoption_prob_mode_2} is deferred to our technical report \cite{TRRRRR} for readability.


    \begin{lem}\label{lem_adoption_prob_mode_1}
      Assume the propagation time of influences follows an arbitrary distribution of p.d.f. $g(x)$. At time step $t$, the probability that $I_1$ reaches the riser first is
      \begin{equation*}
        \frac{X_1(t)}{X_1(t)+X_2(t)}.
      \end{equation*}
    \end{lem}
    \begin{proof}
      Consider a random riser. At time $t$, the expected number of her neighbors infected by $I_1$ (resp. $I_2$) is $X_1(t)p(r,\sigma)$ (resp. $X_2(t)p(r,\sigma)$), which, for ease of exposition, is written as $m$ (resp. $s$).

      Let $Y_1,Y_2,\cdots,Y_m$ (resp. $Z_1,Z_2,\cdots,Z_s$) be the time when the information arrives from users infected by $I_1$ (resp. $I_2$). And r.v. $Y$ (resp. $Z$) denotes the time when the first information about $I_1$ (resp. $I_2$) arrives. Then, $Y=\min\limits_{k=1, \ldots, m}Y_k$ and $Z=\min\limits_{k=1, \ldots, s}Z_k$. User $u$ will adopt $I_1$ if $Y<Z$ and vice versa. Assume the information from neighbors propagate independently. Then, $\{Y_k\}_{k=1}^m$ and $\{Z_k\}_{k=1}^s$ are $i.i.d.$ distributed; $Y$ and $Z$ are independent of each other.

      It is easy to see that the p.d.f. of $Y$ is $f_Y(y)=m[1-G(y)]^{m-1}g(y)$, where $G(\cdot)$ is the cumulative distribution function (c.d.f.) of propagation time. And the p.d.f. of $Z$ is $f_Z(z)=s[1-G(z)]^{s-1}g(z)$. Since $Y$ and $Z$ are independent, we have
      \begin{equation*}
        \begin{aligned}
          P(Y<Z)&=\int_0^{\infty}f_Y(y)\left[\int_y^{\infty}f_Z(z)dz\right]dy  \\
                &=\int_0^{\infty}f_Y(y)\left[\int_y^{\infty}s[1-G(z)]^{s-1}g(z)dz\right]dy  \\
                &=\int_0^{\infty}m[1-G(y)]^{m-1}g(y)[1-G(y)]^sdy  \\
                &=\left.\frac{m}{m+s}[1-G(y)]^{m+s}\right|_{\infty}^0\\
                &=\frac{m}{m+s}.
        \end{aligned}
      \end{equation*}
      Recall that $m=X_1(t)p(r,\sigma)$ and $s=X_2(t)p(r,\sigma)$, then the probability that information about $I_1$ arrives first is $\frac{X_1(t)}{X_1(t)+X_2(t)}$. This completes the proof.
    \end{proof}
    \begin{lem}\label{lem_adoption_prob_mode_2}
      Assume the time for influences to reach a user follows an arbitrary distribution of p.d.f. $g(x)$. At time step $t$, the probability that influence $I_1$ arrives latest is
      \begin{equation*}
        \frac{X_1(t)}{X_1(t)+X_2(t)}.
      \end{equation*}
    \end{lem}

    Regarding strategies (iii) and (iv), we would like to show that they are mathematically equivalent to cases (i) and (ii) respectively. Note that in strategy (iii), users follow the influence of her most similar neighbor, i.e., the one nearest to her. Assume an arbitrary distribution of user distance $g(x)$ in the latent space. Then, by replacing arrival time by the squared Euclidean distance, the setting of (iii) is reduced to case (i). Similarly, assume an arbitrary distribution of user degrees $g(x)$. Substituting the arrival time of strategy (ii) by user's degree, we derive the setting of case (iv).

    Summarizing the four cases, we find that strategies (i) and (iii) could be captured by Lemma \ref{lem_adoption_prob_mode_1}, and strategies (i) and (iv) could be delineated by Lemma \ref{lem_adoption_prob_mode_2}. Particularly, all the four complicated cases implies the same adoption probability.



\subsection{Competing Influences Under Information Overload}

    Recall the diffusion mechanism that under information overload ($t>t_{c}$), a user will adopt influences based on their influence powers with probability $p(t)=e^{-\mu (t-t_{c})}$. And with probability $1-p(t)$, users will resort to the four strategies described in Definition \ref{Def_overloaded_behaviors}. On this basis, similar to the case before overload, we could formulate the behavior of an overloaded riser as follows:
    \begin{itemize}
    \begin{spacing}{1.2}
      \item The new riser will get influenced by $I_1$ with probability \\$p(t)\frac{aX_1(t)}{aX_1(t)+bX_2(t)}+(1-p(t))\frac{X_1(t)}{X_1(t)+X_2(t)}$;
      \item The new riser will get influenced by $I_2$ with probability \\$p(t)\frac{bX_2(t)}{aX_1(t)+bX_2(t)}+(1-p(t))\frac{X_2(t)}{X_1(t)+X_2(t)}$.
    \end{spacing}
    \end{itemize}
    \vspace{-3mm}

    Then, we are ready to formalize the competing process under information overload. Given current states $X_1(t)$ and $X_2(t)$, by applying the continuous analogy technique in Section \ref{sec_before_overload}, we obtain the mean state of the system at time $t+1$
    \begin{equation}\label{eq_differential_eq_after_overload}
      \left\{
          \begin{aligned}
             &\frac{dx_1}{dt}=p(t)\frac{ax_1(t)}{ax_1(t)+bx_2(t)}+(1-p(t))\frac{x_1(t)}{x_1(t)+x_2(t)},\\
             &\frac{dx_2}{dt}=p(t)\frac{bx_2(t)}{ax_1(t)+bx_2(t)}+(1-p(t))\frac{x_2(t)}{x_1(t)+x_2(t)};\\
          \end{aligned}
        \right.
    \end{equation}
    where $x_1(t)$ (resp. $x_2(t)$) is the expectation of $X_1(t)$ (resp. $X_2(t)$) in the continuous analogy.

    In Theorem \ref{Thm_influence_after_overload}, we characterize the influence spread at any time $t>t_{c}$, by solving the above Eq. (\ref{eq_differential_eq_after_overload}).
    \begin{thm}\label{Thm_influence_after_overload}
      The number of users influenced by $I_1$ and $I_2$ satisfies the following equations respectively.

      (i) At time $t_{c}<t<t_{c}+1/\mu$,
      \begin{equation}\label{eq_thm_results_after_overload}
        \left\{
        \begin{aligned}
          &C_3t^{\frac{b-a}{a}\mu t_c}e^{\frac{a-b}{a}\mu t}x_1^{\frac{b}{a}}(t)+x_1(t)=t,\\
          &C_4t^{\frac{a-b}{b}\mu t_c}e^{\frac{b-a}{b}\mu t}x_2^{\frac{a}{b}}(t)+x_2(t)=t,\\
        \end{aligned}
        \right.
      \end{equation}
      where
      \begin{equation}\label{eq_varaiation_thm_after_overload_after_t_c}
        \left\{
          \begin{aligned}
            &C_3=x_2(t_c)/t_c^{\frac{b-a}{a}\mu t_c}e^{\frac{a-b}{a}\mu t_c}x_1^{\frac{b}{a}}(t_c),\\
            &C_4=x_1(t_c)/t_c^{\frac{a-b}{b}\mu t_c}e^{\frac{b-a}{b}\mu t_c}x_2^{\frac{a}{b}}(t_c).\\
          \end{aligned}
        \right.
      \end{equation}

      (ii) At time $t>t_c+1/\mu$,
       \begin{equation}\label{eq_thm_after_overoad_after_t_c_plus}
         x_1(t)=\frac{x_1(t_c+\frac{1}{\mu})}{t_c+\frac{1}{\mu}}t,\; x_2(t)=\frac{x_2(t_c+\frac{1}{\mu})}{t_c+\frac{1}{\mu}}t.
       \end{equation}
    \end{thm}
      \textsc{Proof.}
      Without loss of generality, we focus on the equation about influence $I_1$, i.e.,
      \begin{equation*}
        \frac{dx_1}{dt}=p(t)\frac{ax_1(t)}{ax_1(t)+bx_2(t)}+(1-p(t))\frac{x_1(t)}{x_1(t)+x_2(t)}.
      \end{equation*}
      Note that $x_1(t)+x_2(t)=t$ at any time, the equation is equivalent to
      \begin{equation}\label{eq_after_replace_x_2}
        \frac{dx_1}{dt}=p(t)\frac{ax_1(t)}{(a-b)x_1(t)+bt}+(1-p(t))\frac{x_1(t)}{t}.
      \end{equation}
      Let $q=\frac{x_1(t)}{t}$, then $x_1(t)=qt$ and $\frac{dx_1}{dt}=q+t\frac{dq}{dt}$. After some simplification and separating variables, Eq. (\ref{eq_after_replace_x_2}) becomes
      \begin{equation*}
        \frac{b+(a-b)q}{(a-b)q-(a-b)q^2}dq=\frac{p(t)}{t}dt.
      \end{equation*}
      By splitting the fraction in the left hand side, we have
      \begin{equation*}
        \left(\frac{\frac{b}{a-b}}{q}+\frac{\frac{a}{a-b}}{1-q}\right)dq=\frac{p(t)}{t}dt.
      \end{equation*}
      Integrating both sides and noting that $q<1$, we obtain
      \begin{equation}\label{eq_before_replace_p(t)}
        \frac{b}{a-b}\ln(q)-\frac{a}{a-b}\ln(1-q)=\int\frac{p(t)}{t}dt.
      \end{equation}
      The r.h.s. is an exponential integral and no closed form expression could be derived. To circumvent this problem, we approximate $p(t)$ with its first-order Maclaurin expansion $\hat{p}(t)$,
      \begin{equation*}
      \hat{p}(t)=
        \left\{
          \begin{aligned}
            &-\mu(t-t_c)+1,&&t_c<t<t_c+1/\mu;\\
            &0,&&t\geq t_c+1/\mu.\\
          \end{aligned}
        \right.
      \end{equation*}


      \begin{wrapfigure}{r}{3cm}
        \vspace{-0.5cm}
        \includegraphics[width=3cm]{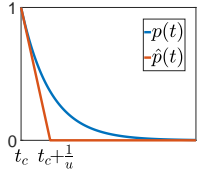}
        \vspace{-0.7cm}
        \caption{$p(t)$ and $\hat{p}(t)$.}\label{fig_p_and_p}
        \vspace{-9mm}
      \end{wrapfigure}
      From Fig. \ref{fig_p_and_p}, we can see that the gap between $\hat{p}(t)$ and $p(t)$ is small, especially when $t$ is around $t_c$ and large enough. Correspondingly, we discuss Eq. (\ref{eq_before_replace_p(t)}) from two parts.

      (i) $t_c<t<t_c+1/\mu$.

      Eq. (\ref{eq_before_replace_p(t)}) is equivalent to
      \begin{eqnarray*} 
      \begin{aligned}
        \frac{b}{a-b}\ln(q)-\frac{a}{a-b}\ln(1-q)
        &=\int -\mu+\frac{1+\mu t_c}{t}dt  \\[-1mm]
        &=\ln e^{-\mu t}+\ln(Ct)^{1+\mu t_c},
      \end{aligned}
      \end{eqnarray*}
      where $C$ is a constant. After some transformation, we have
      \begin{equation*}
        \begin{aligned}
                        &\ln \frac{q^b}{(1-q)^a}=\ln \left[ \frac{(Ct)^{1+\mu t_c}} {e^{\mu t}} \right]^{(a-b)}  \\
         \Longleftrightarrow&q^be^{(a-b)\mu t}=(Ct)^{(1+\mu t_c)(a-b)}(1-q)^a  \\[-1mm]
         \overset{\text{(b)}}{\Longleftrightarrow}&\left(x_1(t)/t\right)^be^{(a-b)\mu t}=(Ct)^{(1+\mu t_c)(a-b)}\left(1\!-\!x_1(t)/t\right)^a\\
         \Longleftrightarrow&x_1^b(t)e^{(a-b)\mu t}=C^{(1+\mu t_c)(a-b)}t^{(a-b)\mu t_c}(t-x_1(t))^a\\
         \Longleftrightarrow&C^{(1+\mu t_c)(b-a)}t^{(b-a)\mu t_c}e^{(a-b)\mu t}x_1^b(t)=(t-x_1(t))^a\\[-1mm]
         \Longleftrightarrow&C^{\frac{(1+\mu t_c)(b-a)}{a}}t^{\frac{b-a}{a}\mu t_c}e^{\frac{a-b}{a}\mu t}x_1^{\frac{b}{a}}(t)=t-x_1(t),\\
        \end{aligned}
      \end{equation*}
      where equality (b) is obtained by recovering $q$ with $\frac{x_1(t)}{t}$. Denoting the constant about $C$ as $C_3$ and solving $C_3$ with the states at $t_c$, we derive the solution of the first case.

      (ii) $t\geq t_c+1/\mu$.

      In this case, Eq. (\ref{eq_before_replace_p(t)}) reduces to
      \begin{equation*}
        \frac{dx_1}{dt}=\frac{x_1(t)}{x_1(t)+x_2(t)},
      \end{equation*}
      which coincides with a special case of Eq. (\ref{eq_differential_eq_before_overload}), where $a=b$. By Corollary \ref{cor_specific_expression}, we obtain that
      \begin{equation*}
        x_1(t)=\frac{x_1(t_c+1/\mu)}{t_c+1/\mu}t, \;\;x_2(t)=\frac{x_2(t_c+1/\mu)}{t_c+1/\mu}t
      \end{equation*}

      Combining cases (i) and (ii), we complete the proof.
    \qed

      \textbf{Remark.} Theorem \ref{Thm_influence_after_overload} reveals that under information overload, the competing dynamics are different from before. Assume $a>b$, for $t_c<t<t_c+1/\mu$, similar to the derivation of Eq. (\ref{eq_variation_thm_results_before_overload}), from Eq. (\ref{eq_thm_results_after_overload}) we could obtain
      \begin{equation}\label{eq_variation_thm_results_after_overload}
      \setlength{\abovedisplayskip}{3pt}
            x_1(t)=t/[1+C_3t^{\frac{b-a}{a}\mu t_c}e^{\frac{a-b}{a}\mu t} x_1^{\frac{b}{a}-1}(t)],
      \setlength{\abovedisplayskip}{3pt}
        \end{equation}
      where the term before $x_1^{\frac{b}{a}-1}(t)$ is significantly larger than that of Eq. (\ref{eq_variation_thm_results_before_overload}), implying that if given the same initial states, $x_1(t)$ would be smaller than that before overload, i.e., the stronger influence is impaired after overload. After time $t_c$, applying similar analyses to Eq. (\ref{eq_thm_after_overoad_after_t_c_plus}), we could see that the impairment is even severer. Specifically, the stronger influence stops increasing and becomes stabilizing, let alone ``taking all''.

    Finally, we make a rigorous comparison between the competing dynamics before and after information overload by Theorem \ref{thm_compare_before_and after_overload}. To this end, we treat the two cases as individuals instead of consecutive sub-processes and assume the same initial states, i.e., $x_1(t_0)=x_1^\prime(t_0)$ and $x_2(t_0)=x_2^\prime(t_0)$, where $x_{1,2}(\cdot)$ (resp. $x_{1,2}^\prime(\cdot)$) denotes the influences before (resp. after) information overload (correspondingly referred to as case 1 and case 2 respectively for brevity).

    \begin{thm}\label{thm_compare_before_and after_overload}
      At ant time $t$, the winner's influence in case 2 is no larger than that in case 1 and the weaker's influence in case 2 is no smaller than that in case 1, i.e.,
      \begin{subequations}
      \begin{numcases}{}
      x_1^\prime(t)\leq x_1(t) \;\text{ and } \;x_2^\prime(t)\geq x_2(t),\;\;\;\;\;\;\text{if } a>b;\;\;\;\;\;\;\label{eq_compare_before_and_after_overload_a}\\
      x_1^\prime(t)\geq x_1(t) \;\text{ and } \;x_2^\prime(t)\leq x_2(t),\;\;\;\;\;\;\text{if } a<b.\;\;\;\;\;\;\label{eq_compare_before_and_after_overload_b}
      \end{numcases}
      \end{subequations}
    \end{thm}
    \begin{proof}
      We focus on proving the results under $a>b$, i.e., $I_1$ being the winner. And the results under $a<b$ could be proven in a similar way.

      To prove the inequality in Eq. (\ref{eq_compare_before_and_after_overload_a}), for $x_1(t)$, by Eq. (\ref{eq_variation_thm_results_before_overload}) we have
      \begin{equation}\label{eq_recall_of_variation_x_1}
        x_1(t)=\frac{t}{1+C_1x_1^{\frac{b}{a}-1}(t)},
      \end{equation}
      where $C_1$ is $x_2(t_0)/x_1^{\frac{b}{a}}(t_0)$.
      For $x_1^\prime(t)$, recall the results in Theorem \ref{Thm_influence_after_overload}, we could transform it to 
      \begin{subequations}
      \begin{numcases}{x_1^\prime(t)=}
      \frac{t}{1\!+\!C_3^\prime [x_1^\prime(t)]^{\frac{b}{a}-1}},\;\;t_0<t<t_0+1/\mu\;\;\;\;\label{eq_variation_of_thm3_a}\\
            t\cdot x_1^\prime(t_0+\frac{1}{\mu})/ (t_0+\frac{1}{\mu}).\;\;\;t\geq t_0+\frac{1}{\mu}\;\;\;\;\label{eq_variation_of_thm3_b}
      \end{numcases}
      \end{subequations}
      where $C_3^\prime=x_2^\prime(t_0)t^{\frac{b-a}{a}\mu t_0}e^{\frac{a-b}{a}\mu t}/t_0^{\frac{b-a}{a}\mu t_0}e^{\frac{a-b}{a}\mu t_0}[x_1^\prime(t_0)]^{\frac{b}{a}}$. Denoting $\rho(t)=t^{\frac{b-a}{a}\mu t_0}e^{\frac{a-b}{a}\mu t}/t_0^{\frac{b-a}{a}\mu t_0}e^{\frac{a-b}{a}\mu t_0}$, we could write $C_3^\prime=C_1\rho(t)$ since $x_1^\prime(t_0)=x_1(t_0)$ and $x_2^\prime(t_0)=x_2(t_0)$.

      Seeing that $x_1^\prime(t)$ is a piece-wise function, we compare $x_1(t)$ with the two cases of $x_1^\prime(t)$ respectively.

      (i) $t_0<t<t_0+1/\mu$

      In this condition, the main difference between $x_1(t)$ and $x_1^\prime(t)$ lies in the time varying coefficient $\rho(t)$. To examine the property of $\rho(t)$, for $t\in [t_0, t_0+1/\mu]$, we write $t$ as $t_0+\Delta$ where $\Delta$ is a variable ranging in $[0,\frac{1}{\mu}]$. Then, we have
      \begin{equation}\label{eq_expansion_of_rho}
      \begin{aligned}
         \rho(t)=&\rho(\Delta,t_0) = \frac{ (t_0+\Delta)^{\frac{b-a}{a}\mu t_0}e^{\frac{a-b}{a}\mu (t_0+\Delta)}  }{  t_0^{\frac{b-a}{a}\mu t_0}e^{\frac{a-b}{a}\mu t_0}  }\\[-1mm]
         =&\,e^{\frac{a-b}{a}\mu \Delta} \left(1+\frac{\Delta}{t_0}\right)^{ \frac{b-a}{a}\mu t_0 }
         =\,\left[e/\left(1+\frac{\Delta}{t_0}\right)^{\frac{t_0}{\Delta}}\right]^{\frac{a-b}{a}\mu \Delta}.
      \end{aligned}
    \end{equation}
    Since $(1+x)^{\frac{1}{x}}$ is monotonically decreasing, we have $\left(1+\frac{\Delta}{t_0}\right)^{\frac{t_0}{\Delta}}\leq \lim\limits_{x\rightarrow 0}(1+x)^{\frac{1}{x}}=e$. Recall that $a>b$, then $\rho_1(t)\geq1$ and is monotonically increasing over time $t$.

    Combining Eq.s (\ref{eq_recall_of_variation_x_1}) and (\ref{eq_variation_of_thm3_a}) and noting that $C_3^\prime=C_1\rho(t)$, we could derive the equation below
    \begin{equation}\label{eq_equality_of_I}
      \vspace{-1mm}
      x_1(t)+C_1x_1^{\frac{b}{a}}(t)=x_1^\prime(t)+C_1\rho(t)[x_1^\prime(t)]^{\frac{b}{a}}.
      \vspace{-1mm}
    \end{equation}
    The above equation (\ref{eq_equality_of_I}) is enough for us to draw our conclusion that $x_1^\prime(t)\leq x_1(t)$. To see this, by contradiction, we assume $x_1^\prime(t)> x_1(t)$. Then, we have
    \begin{equation*}
      x_1^\prime(t)+C_1\rho(t)[x_1^\prime(t)]^{\frac{b}{a}}>x_1(t)+C_1\rho(t)[x_1(t)]^{\frac{b}{a}}.
    \end{equation*}
    Recall Eq. (\ref{eq_equality_of_I}), we further derive
    \begin{equation*}
      x_1(t)+C_1[x_1(t)]^{\frac{b}{a}}>x_1(t)+C_1\rho(t)[x_1(t)]^{\frac{b}{a}},
    \end{equation*}
    which implies $\rho(t)<1$, contradicting with our finding $\rho(t)\geq 1$. Therefore, $x_1^\prime(t)\leq x_1(t)$.

    Note that $x_1^\prime(t)+x_2^\prime(t)=x_1(t)+x_2(t)=t$. By $x_1^\prime(t)\leq x_1(t)$, we could derive that $x_2^\prime(t)\geq x_2(t)$.


    (ii) $t\geq t_0+1/\mu$

    For easy of notation, we denote $t_0+1/\mu$ as $t_0^\prime$ and consider the influence diffusion thereafter, i.e., treating $t_0^\prime$ as the new start with states consistent with $t\in (t_0, t_0+1/\mu)$. Then, the influence spread of $I_1$ could be expressed as
      \begin{equation}\label{eq_variation_x_1_bigger_t}
        \vspace{-1mm}
        x_1(t)=t/[1+\widehat{C_1} x_1^{\frac{b}{a}-1}(t)],\;\;\;\;t\geq t_0+1/\mu,
        \vspace{-1mm}
      \end{equation}
    where $\widehat{C_1}=x_2(t_0^\prime)/x_1^{\frac{b}{a}}(t_0^\prime)$.

    Then, combining Eq.s (\ref{eq_variation_x_1_bigger_t}) and (\ref{eq_variation_of_thm3_b}) and noting that $t_0^\prime=x_1^\prime(t_0^\prime)+x_2^\prime(t_0^\prime)$, we have
    \begin{equation}\label{eq_equality_I_bigger_t}
      x_1(t)+\frac{x_2(t_0^\prime)  } {  x_1^{\frac{b}{a}}(t_0^\prime)}  x_1^{\frac{b}{a}}(t)=\frac{x_1^\prime(t)t_0^\prime}{x_1^\prime(t_0^\prime)}=x_1^\prime(t) + \frac{x_2^\prime(t_0^\prime)  } {  x_1^\prime(t_0^\prime)}x_1^\prime(t).
    \end{equation}
    Again, we tend to prove $x_1^\prime(t)\leq x_1(t)$ by contradiction. To this end, we assume $x_1^\prime(t)> x_1(t)$. Then, Eq. (\ref{eq_equality_I_bigger_t}) implies that
    \begin{equation}\label{eq_contradiction_for_bigger_t}
    \begin{aligned}
      &x_1(t)\;\;+\!\!\!&\frac{x_2(t_0^\prime)  } {  x_1^{\frac{b}{a}}(t_0^\prime)}  x_1^{\frac{b}{a}}(t)  &>  x_1(t) + \frac{x_2^\prime(t_0^\prime)  } {  x_1^\prime(t_0^\prime)}x_1(t)\\
      \Leftrightarrow& &\frac{x_2(t_0^\prime)  }  {  x_2^\prime(t_0^\prime)  } \frac{  x_1^\prime(t_0^\prime)}  {  x_1^{\frac{b}{a}}(t_0^\prime)}   &> [x_1(t)]^{1-\frac{b}{a}}\\[-1mm]
      \overset{\text{(c)}}{\Leftrightarrow}& &[x_1(t_0^\prime)]^{1-\frac{b}{a}}  &>  [x_1(t)]^{1-\frac{b}{a}}\\[-2mm]
      \overset{\text{(d)}}{\Leftrightarrow}& &x_1(t_0^\prime)  &>  x_1(t),
    \end{aligned}
    \end{equation}
    where step (c) is due to our results under $t_0<t<t_0+1/\mu$ that $x_1^\prime(t_0^\prime)\leq x_1(t_0^\prime)$ and $x_2^\prime(t_0^\prime)\geq x_2(t_0^\prime)$, and step (d) is due to the premise $a>b$. Note that $x_1(t)$ is non-decreasing over time and $t\geq t_0^\prime$. Thus the conclusion in Eq. (\ref{eq_contradiction_for_bigger_t}) contradicts with the monotonicity of $x_1(t)$. And we have $x_1^\prime(t)\leq x_1(t)$ for $t\geq t_0+1/\mu$.

    Following similar technique in case (i), we derive $x_2^\prime(t)\geq x_2(t)$, since $x_1^\prime(t)+x_2^\prime(t)=x_1(t)+x_2(t)=t$.

    \vspace{1mm}
    Combining cases (i) and (ii), we finish the proof of Eq. (\ref{eq_compare_before_and_after_overload_a}). To obtain the symmetric result Eq. (\ref{eq_compare_before_and_after_overload_b}), we only need to swap $a$ and $b$ and thus the detailed derivations are omitted.
    \end{proof}

    \textbf{Remark.} Theorem \ref{thm_compare_before_and after_overload} indicates that in the case of overload, the advantage of the stronger influence is impaired and the weaker influence could gain more followers, resulting in a smaller difference between $I_1$ and $I_2$. That is, the competition after overload seems to be milder than that before overload.
    
    Synthesizing Theorems \ref{Thm_influence_before_overload}, \ref{Thm_influence_after_overload} and \ref{thm_compare_before_and after_overload}, we see that in the beginning, the share of the stronger influence constantly increases and the weaker one decreases. However, with the advent of overload, the competition becomes milder and after a short time ($1/\mu$), both influences tend to stabilize, as if the competition is over.
    

\section{Discussion}\label{sec_discussion}

    To answer the question posed in the title and illuminate the implications of our theoretical findings, we present a detailed discussion on the following issues.

    \textbf{Question:} \underline{\textit{Will the winner take all?}}

    \textbf{Answer:} The winner could take all under certain condition but not always. The result is a joint effect of the intrinsic, subjective and objective factors. Generally speaking, if users are not to be overloaded, the stronger influence will win out and dominate the network, exhibiting the ``winner takes all'' phenomenon. In the case of information overload, the result depends on the stable state, which roughly begins at the time of overload and keeps steady till the final state.

    \textbf{Question:} \underline{\textit{What determines the final states?}}

    \textbf{Answer:} Before overload, the final state is solely determined by the influence power. The influence with a larger power will dominate nearly the whole network. In the case of overload, the final state is jointly determined by the initial state, influence power and the time of overload. Specifically, to win the competition, an influence is expected to have more initial followers, larger influence power and a later advent of overload, to establish and enhance its advantage before the advent of overload.

    Further, the overload is likely to arrive late if the network is sparsely connected. And users are robust to superfluous information if they are less active in social interactions.

    \textbf{Question:} \ul{\textit{Are the theoretical results of practical significance in events like COVID-19?}}

    \textbf{Answer:} Absolutely. Take the ``sesame treatment'' as an example. Suppose the rumor has more initial followers and larger influence power, due to its broadcast on media and users' panic for new treatments. In turn, the truth is situated in a negative condition. By our findings, to win the competition, we should first publish the truth timely to earn more initial followers. Secondly, we could attach scientific evidences, advice of the authority etc. to the propagation of the truth, as a means of elevating its influence power.

    Thirdly, the use of information overload is a little tricky. Specifically, when the share of the truth is smaller, the overload is desired to restrain the false, and vice versa.
    To harness the overload, we could adjust the network connectivity by blocking or recommending links; control the influence range by promoting or restricting user's activeness; affect user's sense of overload by flooding them with information or filtering information for them.

\section{Experiments}\label{sec_experiment}

  In this section, we validate our theoretical findings by simulating influence competitions on real-world social networks.
\subsection{Dataset Description}

  Two datasets are included for experiments, with detailed information as follows. As can be seen, the two networks are of nearly the same size of users, while BlogCatalog is shown to have 12 times larger amount of edges than Ca-HepTH, implying much denser user connections, i.e., smaller $\sigma^2$.

  $\bullet$ \textbf{BlogCatalog} is a social blog directory website, managing the bloggers and their blogs \cite{BlogCatalog}. This dataset contains 10,312 users and 333,983 edges.

  $\bullet$ \textbf{Ca-HepTh} is a collaboration network with 9,877 users and 25,998 edges, covering collaborations between authors who submitted papers to the HepTh category in arXiv \cite{dataset_ca_hepth}.

\begin{figure}[h]
    \centering
      \subfigure[BlogCatalog]
        {
        \begin{minipage}[h]{0.232\textwidth}
        \vspace{-3.7mm}
            \includegraphics[width=1\textwidth]{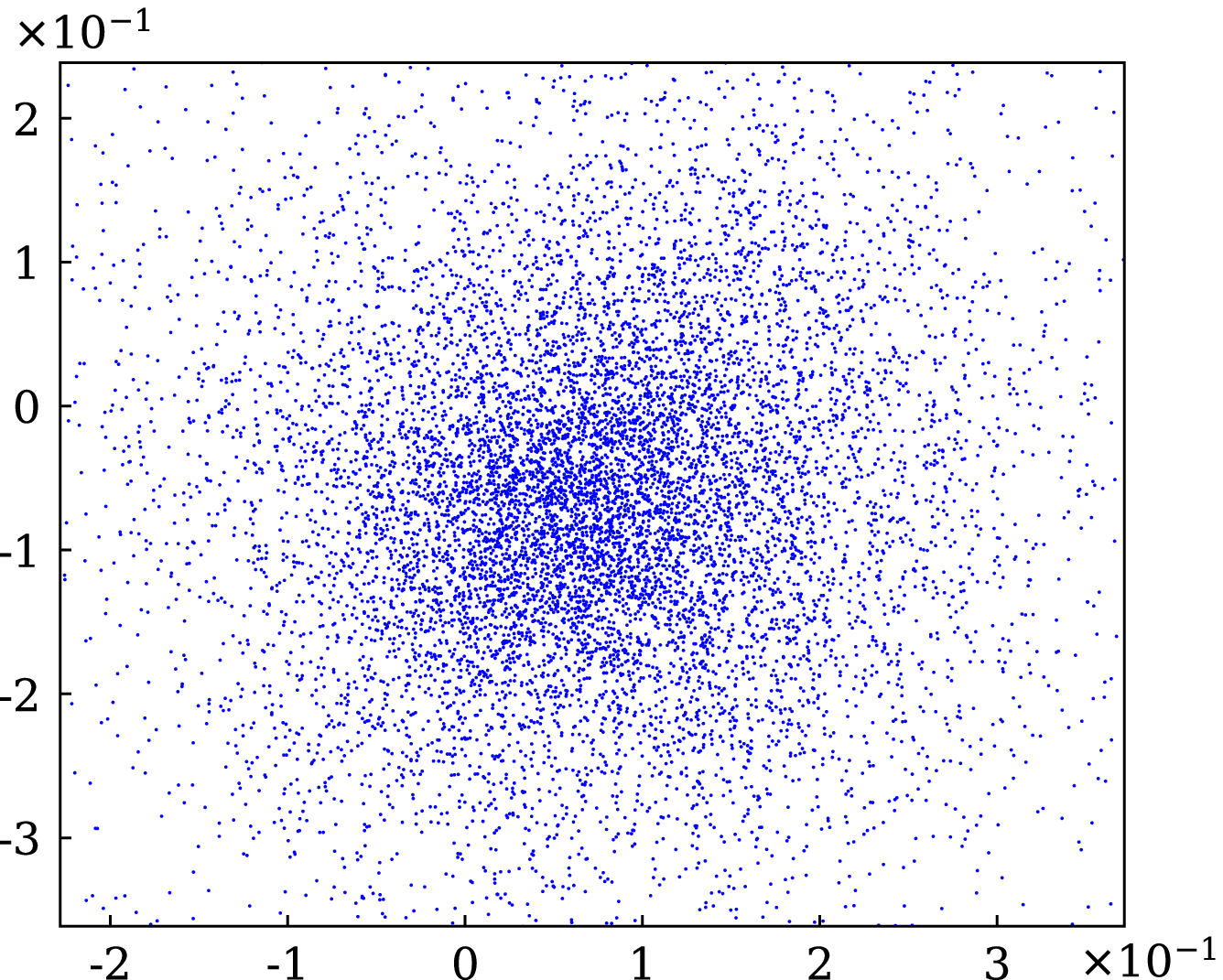}\label{fig_BC_emb}
            \vspace{-3.2mm}
          \end{minipage}}
      \hspace{-2mm}  
      \subfigure[Ca-HepTh]
      {
      \begin{minipage}[h]{0.232\textwidth}
        \vspace{-3.7mm}
           \includegraphics[width=1\textwidth]{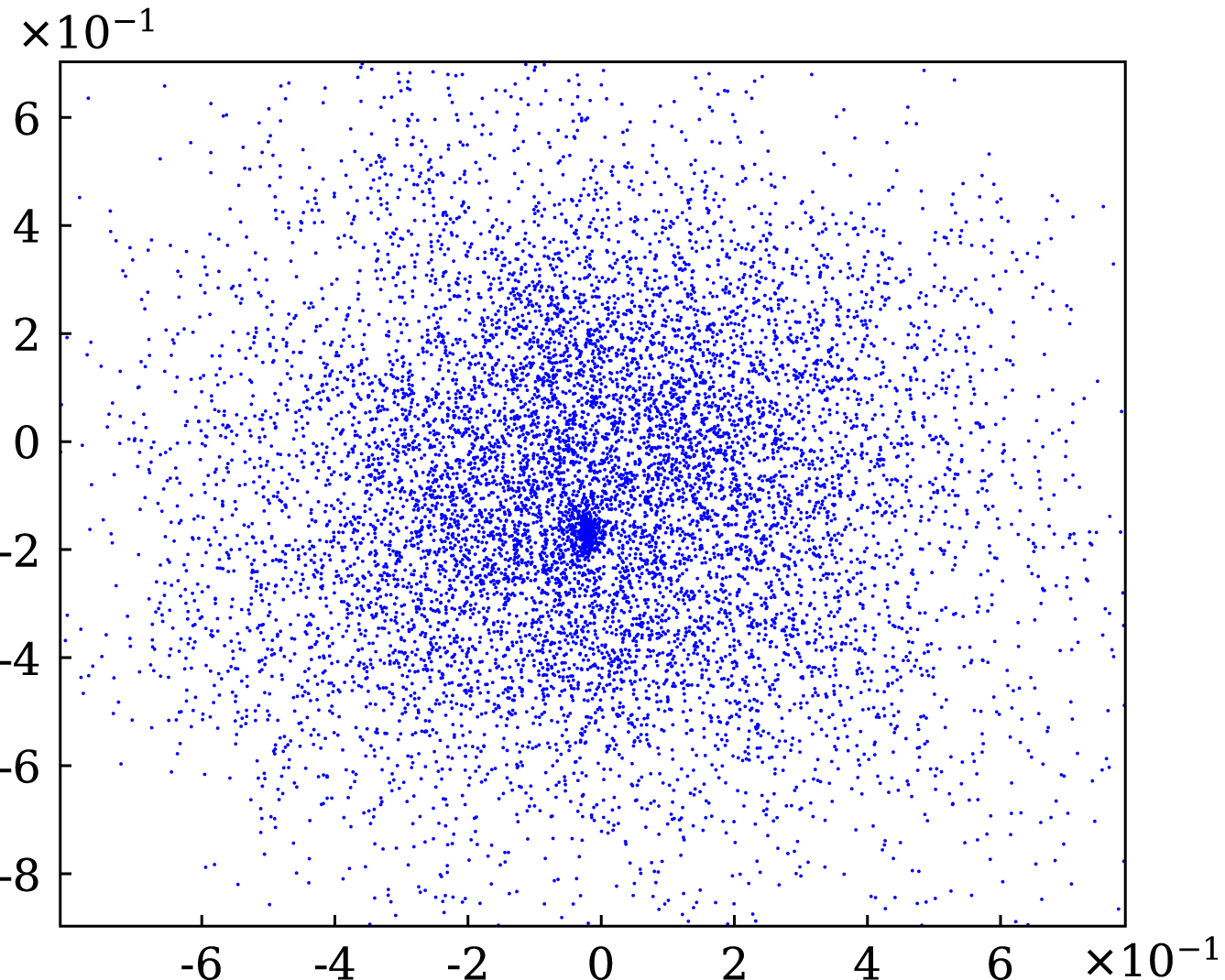}\label{fig_CA_emb}
           \vspace{-3.2mm}
         \end{minipage}}
      \vspace{-4mm}
      \caption{Visualization of the embedded networks.}\label{fig_embed_visual}
      \vspace{-6mm}
    \end{figure}

    In preparation for running competing processes, we first convert the networks into a latent space by the ``node2vec'' technique, where the user distance is instantiated as the squared Euclidean distance and the parameters are set by default (esp. $d=128$). For an intuitive vision, we visualize user embeddings of two random dimensions in Fig. \ref{fig_embed_visual}. It is shown that the embeddings of BlogCatalog are evidently denser than that of Ca-HepTh, due to the gap in the density of edges. This observation is also verified by the estimation of the variance $\sigma^2$, which is found to be 0.0147 and 0.0916 respectively.


\subsection{Experimental Settings}

  Basically, we set influence range $r$ to be 4 and $\mu$ to be 10.
  Before information overload, for effective validation, we assume $t_c$ is large enough by setting $\delta_c=|V|$. Two pairs of initial states are considered: 40\% $I_1$  versus 60\% $I_2$ (denoted as S1) and the reverse case 60\% $I_1$  versus 40\% $I_2$ (denoted as S2), with total initial followers $X_1(t_0)+X_2(t_0)=40$. Further, $I_1$ is selected to be the weaker one with influence power $a=1$. And $I_2$ is tested under power $b=2$ and 3 for comparison.

  In the case of overload, the influence powers are set to be $a=1$ and $b=2$. To test the effect of overload timing, the competition is run under both $\delta_c=30$ and $\delta_c=500$. Moreover, to demonstrate that the weaker influence is possible to be the winner, besides S1 we consider a new case S3 (80\% $I_1$ versus 20\% $I_2$) where $I_1$ is given a larger initial advantage.

\begin{figure*}[t]
\centering
\includegraphics[width=1\textwidth]{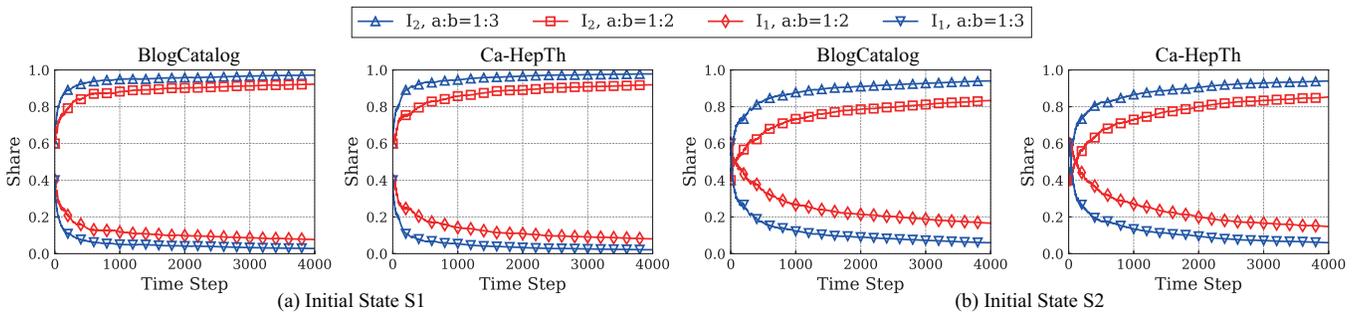}
\vspace{-0.6cm}
\caption{Competing dynamics BEFORE information overload.}
\label{fig_before_overload}
\vspace{-2mm}
\end{figure*}

\begin{figure*}[t]
\centering
\includegraphics[width=1\textwidth]{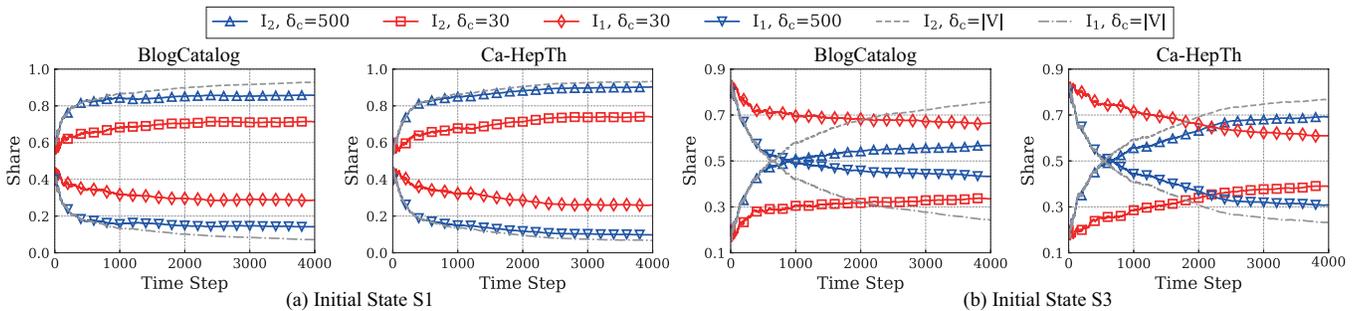}
\vspace{-0.6cm}
\caption{Competing dynamics AFTER information overload.}
\label{fig_after_overload}
\vspace{-3mm}
\end{figure*}

\subsection{Experimental Results}

  The competing dynamics, featured by the share of each influence, are reported in Figs. \ref{fig_before_overload} and \ref{fig_after_overload}, corresponding to cases before and after information overload respectively.

  \subsubsection{\textbf{Before Overload}}
  For both initial states S1 and S2 in Fig. \ref{fig_before_overload}, the stronger influence (i.e., $I_2$) is shown to be the winner and gradually dominate the network, verifying the ``winner takes all'' phenomenon revealed by Theorem \ref{Thm_influence_before_overload}.
  
  Specifically, under initial state S1, as one may imagine, with a greater influence power and more initial followers, $I_2$ undoubtedly becomes the winner and quickly dominates nearly the whole network, as shown in Fig. \ref{fig_before_overload}(a). More precisely, it is observed that the line marked by ``$\vartriangle$'' ($I_2$, $b=3$) grows faster than that marked ``$\square$'' ($I_2$, $b=2$), since $I_2$ gains more advantage over $I_1$ when $b=3$, as predicted by Theorem \ref{Thm_influence_before_overload}.


  Under initial state S2, although $I_2$ is given less initial followers than $I_1$, as can be seen in Fig. \ref{fig_before_overload}(b), $I_2$ still managed to win the competition with a share approaching 100\%. This observation serves as a firm validation of Theorem \ref{Thm_influence_before_overload} which implies that the stronger influence will win nearly the whole network anyway, regardless of the initial condition.
  
  \subsubsection{\textbf{After Overload}}

  The competing process after information overload evidently differs from that before overload, exhibiting more complex competing dynamics. 
  
  The results under initial state S1 are reported in Fig. \ref{fig_after_overload}(a). It is shown that even with a greater influence power and more initial followers, $I_2$ is not assured to dominate the whole network as it could have done without overload (the dashed grey line), and only maintains a stable share of market, as predicted by Theorem \ref{Thm_influence_after_overload}. Comparing the lines marked ``$\vartriangle$'' ($I_2$, $\delta_c=500$) and ``$\square$'' ($I_2$, $\delta_c=30$), we see that the final share of $I_2$ is increased when $\delta_c=500$, since by Theorem \ref{Thm_overload_time} the advent of overload is postponed with a larger $\delta_c$ and therefore $I_2$ is given more time to increase. Moreover, we find that the share of the stronger influence without overload (the dashed grey line) is larger  than that under overload (lines marked ``$\vartriangle$'' and ``$\square$''), thus verifying the results in Theorem \ref{thm_compare_before_and after_overload}.
  
   We further simulate the competition under initial state S3 to show that the stronger influence $I_2$ is not assured to be the winner, let alone ``taking all''. Instead, the weaker one $I_1$ is possible to win the competition. As can be seen in Fig. \ref{fig_after_overload}(b), starting from a larger initial share of 80\%, $I_1$ is constantly decreasing in the beginning. However, with a small $\delta_c=30$ (i.e., early advent of overload), the tendency of decline is stopped timely before losing its advantage (Theorem \ref{Thm_influence_after_overload}). Resultantly, $I_1$ becomes the final winner though with a weaker power. Conversely, under $\delta_c=500$, the time of overload is postponed (Theorem \ref{Thm_overload_time}). Thus, the share of $I_1$ continues to decrease and becomes inferior to $I_2$ finally.

  Comparing the results under an early and late advent of overload, we could infer that given the same initial states, the weaker influence in $G(V,E^\prime)$ would have a larger share than that in $G(V,E)$, since $|E^\prime|\geq |E|$ due to the link recovery which implies a smaller $\sigma^2$ and an early advent of overload.

  By Figures \ref{fig_before_overload} and \ref{fig_after_overload}, we validate the theoretical findings which are found to be in good agreement with the experimental results.


\section{Conclusion}\label{sec_conclusion}

    In this paper, we present a comprehensive study into the competing dynamics in social networks. Different from existing work, we considered not only the intrinsic factor but also the subjective and objective factors. Specifically, the network is mapped into the latent space to recover possible diffusion links. And users are allowed to adopt different information strategies when overloaded. In this context, we formulate the cases before and after overload. Finally, explicit expressions of competing dynamics are derived. It is found that before information overload, the competing process exhibits the ``winner takes all'' phenomenon. While in the case of overload, the final state is determined by the stable state which is a joint result of the considered factors.

\footnotesize
\bibliographystyle{ACM-Reference-Format}
\bibliography{ref_opinion_battle}


\begin{thebibliography}{24}


\ifx \showCODEN    \undefined \def \showCODEN     #1{\unskip}     \fi
\ifx \showDOI      \undefined \def \showDOI       #1{#1}\fi
\ifx \showISBNx    \undefined \def \showISBNx     #1{\unskip}     \fi
\ifx \showISBNxiii \undefined \def \showISBNxiii  #1{\unskip}     \fi
\ifx \showISSN     \undefined \def \showISSN      #1{\unskip}     \fi
\ifx \showLCCN     \undefined \def \showLCCN      #1{\unskip}     \fi
\ifx \shownote     \undefined \def \shownote      #1{#1}          \fi
\ifx \showarticletitle \undefined \def \showarticletitle #1{#1}   \fi
\ifx \showURL      \undefined \def \showURL       {\relax}        \fi
\providecommand\bibfield[2]{#2}
\providecommand\bibinfo[2]{#2}
\providecommand\natexlab[1]{#1}
\providecommand\showeprint[2][]{arXiv:#2}

\bibitem[\protect\citeauthoryear{Barab{\'a}si and Albert}{Barab{\'a}si and
  Albert}{1999}]%
        {nature_power_law}
\bibfield{author}{\bibinfo{person}{Albert-L{\'a}szl{\'o} Barab{\'a}si} {and}
  \bibinfo{person}{R{\'e}ka Albert}.} \bibinfo{year}{1999}\natexlab{}.
\newblock \showarticletitle{Emergence of scaling in random networks}.
\newblock \bibinfo{journal}{\emph{science}} \bibinfo{volume}{286},
  \bibinfo{number}{5439} (\bibinfo{year}{1999}), \bibinfo{pages}{509--512}.
\newblock


\bibitem[\protect\citeauthoryear{De~Choudhury, Sundaram, John, Seligmann, and
  Kelliher}{De~Choudhury et~al\mbox{.}}{2010}]%
        {birds_2}
\bibfield{author}{\bibinfo{person}{Munmun De~Choudhury}, \bibinfo{person}{Hari
  Sundaram}, \bibinfo{person}{Ajita John}, \bibinfo{person}{Doree~Duncan
  Seligmann}, {and} \bibinfo{person}{Aisling Kelliher}.}
  \bibinfo{year}{2010}\natexlab{}.
\newblock \showarticletitle{`Birds of a Feather': Does User Homophily Impact
  Information Diffusion in Social Media?}
\newblock \bibinfo{journal}{\emph{arXiv preprint arXiv:1006.1702}}
  (\bibinfo{year}{2010}).
\newblock


\bibitem[\protect\citeauthoryear{Feng, Sun, and Fu}{Feng et~al\mbox{.}}{2020}]%
        {TRRRRR}
\bibfield{author}{\bibinfo{person}{Chen Feng}, \bibinfo{person}{Jiahui Sun},
  {and} \bibinfo{person}{Luoyi Fu}.} \bibinfo{year}{2020}\natexlab{}.
\newblock \bibinfo{title}{Will the Winner Take All? Competing Influences in
  Social Networks Under Information Overload}.
\newblock
  \bibinfo{howpublished}{\url{https://www.dropbox.com/s/hlhhq69rz402393/OB-mobihoc.pdf?dl=0}}.
\newblock


\bibitem[\protect\citeauthoryear{{Google Trends}}{{Google Trends}}{2020}]%
        {google_trends}
\bibfield{author}{\bibinfo{person}{{Google Trends}}.} \bibinfo{year}{(accessed
  December 9, 2020)}\natexlab{}.
\newblock \bibinfo{title}{Search count of {Myspace} and {Facebook}, {iOS} and
  {Android}}.
\newblock \bibinfo{howpublished}{https://trends.google.com/trends/}.
\newblock


\bibitem[\protect\citeauthoryear{Grover and Leskovec}{Grover and
  Leskovec}{2016}]%
        {node2vec}
\bibfield{author}{\bibinfo{person}{Aditya Grover} {and} \bibinfo{person}{Jure
  Leskovec}.} \bibinfo{year}{2016}\natexlab{}.
\newblock \showarticletitle{node2vec: Scalable feature learning for networks}.
  In \bibinfo{booktitle}{\emph{Proc. Int. Conf. Knowl. Discov. Data Mining,
  KDD.}} ACM, \bibinfo{pages}{855--864}.
\newblock


\bibitem[\protect\citeauthoryear{Jacoby}{Jacoby}{1984}]%
        {jacob_overload}
\bibfield{author}{\bibinfo{person}{Jacob Jacoby}.}
  \bibinfo{year}{1984}\natexlab{}.
\newblock \showarticletitle{Perspectives on information overload}.
\newblock \bibinfo{journal}{\emph{Journal of consumer research}}
  \bibinfo{volume}{10}, \bibinfo{number}{4} (\bibinfo{year}{1984}),
  \bibinfo{pages}{432--435}.
\newblock


\bibitem[\protect\citeauthoryear{Koprulu, Kim, and Shroff}{Koprulu
  et~al\mbox{.}}{2019}]%
        {opinion_battle}
\bibfield{author}{\bibinfo{person}{Irem Koprulu}, \bibinfo{person}{Yoora Kim},
  {and} \bibinfo{person}{Ness~B Shroff}.} \bibinfo{year}{2019}\natexlab{}.
\newblock \showarticletitle{Battle of opinions over evolving social networks}.
\newblock \bibinfo{journal}{\emph{{IEEE/ACM} Trans. Netw.}}
  \bibinfo{volume}{27}, \bibinfo{number}{2} (\bibinfo{year}{2019}),
  \bibinfo{pages}{532--545}.
\newblock


\bibitem[\protect\citeauthoryear{Leskovec, Kleinberg, and Faloutsos}{Leskovec
  et~al\mbox{.}}{2007}]%
        {dataset_ca_hepth}
\bibfield{author}{\bibinfo{person}{Jure Leskovec}, \bibinfo{person}{Jon
  Kleinberg}, {and} \bibinfo{person}{Christos Faloutsos}.}
  \bibinfo{year}{2007}\natexlab{}.
\newblock \showarticletitle{Graph evolution: Densification and shrinking
  diameters}.
\newblock \bibinfo{journal}{\emph{{ACM} Trans. Knowl. Discovery from Data}}
  \bibinfo{volume}{1}, \bibinfo{number}{1} (\bibinfo{year}{2007}),
  \bibinfo{pages}{2--es}.
\newblock


\bibitem[\protect\citeauthoryear{Liao, He, Zhang, and Chua}{Liao
  et~al\mbox{.}}{2018}]%
        {Embedding_link_prediction_1}
\bibfield{author}{\bibinfo{person}{Lizi Liao}, \bibinfo{person}{Xiangnan He},
  \bibinfo{person}{Hanwang Zhang}, {and} \bibinfo{person}{Tat-Seng Chua}.}
  \bibinfo{year}{2018}\natexlab{}.
\newblock \showarticletitle{Attributed social network embedding}.
\newblock \bibinfo{journal}{\emph{{IEEE} Trans. Knowl. Data Eng.}}
  \bibinfo{volume}{30}, \bibinfo{number}{12} (\bibinfo{year}{2018}),
  \bibinfo{pages}{2257--2270}.
\newblock


\bibitem[\protect\citeauthoryear{Liu, Par{\'e}, Nedich, Tang, Beck, and
  Basar}{Liu et~al\mbox{.}}{2019}]%
        {Basar_Tamer_bi_virus}
\bibfield{author}{\bibinfo{person}{Ji Liu}, \bibinfo{person}{Philip~E
  Par{\'e}}, \bibinfo{person}{Angelia Nedich}, \bibinfo{person}{Choon~Yik
  Tang}, \bibinfo{person}{Carolyn~L Beck}, {and} \bibinfo{person}{Tamer
  Basar}.} \bibinfo{year}{2019}\natexlab{}.
\newblock \showarticletitle{Analysis and control of a continuous-time bi-virus
  model}.
\newblock \bibinfo{journal}{\emph{IEEE Trans. Automat. Contr.}}
  (\bibinfo{year}{2019}).
\newblock


\bibitem[\protect\citeauthoryear{McPherson, Smith-Lovin, and Cook}{McPherson
  et~al\mbox{.}}{2001}]%
        {birds_of_a_feather}
\bibfield{author}{\bibinfo{person}{Miller McPherson}, \bibinfo{person}{Lynn
  Smith-Lovin}, {and} \bibinfo{person}{James~M Cook}.}
  \bibinfo{year}{2001}\natexlab{}.
\newblock \showarticletitle{Birds of a feather: Homophily in social networks}.
\newblock \bibinfo{journal}{\emph{Ann. Rev. Sociol.}} \bibinfo{volume}{27},
  \bibinfo{number}{1} (\bibinfo{year}{2001}), \bibinfo{pages}{415--444}.
\newblock


\bibitem[\protect\citeauthoryear{Muchnik, Pei, Parra, Reis, Andrade~Jr, Havlin,
  and Makse}{Muchnik et~al\mbox{.}}{2013}]%
        {power_law_2}
\bibfield{author}{\bibinfo{person}{Lev Muchnik}, \bibinfo{person}{Sen Pei},
  \bibinfo{person}{Lucas Parra}, \bibinfo{person}{Saulo Reis},
  \bibinfo{person}{Jos{\'e}~S Andrade~Jr}, \bibinfo{person}{Shlomo Havlin},
  {and} \bibinfo{person}{Hern{\'a}n~A Makse}.} \bibinfo{year}{2013}\natexlab{}.
\newblock \showarticletitle{Origins of power-law degree distribution in the
  heterogeneity of human activity in social networks}.
\newblock \bibinfo{journal}{\emph{Sci. Rep.}} \bibinfo{volume}{3},
  \bibinfo{number}{1} (\bibinfo{year}{2013}), \bibinfo{pages}{1--8}.
\newblock


\bibitem[\protect\citeauthoryear{Osaki}{Osaki}{2020}]%
        {Sesame_oil}
\bibfield{author}{\bibinfo{person}{Tomohiro Osaki}.} \bibinfo{year}{(accessed
  Dec. 9, 2020)}\natexlab{}.
\newblock \bibinfo{title}{Sesame oil? Granite baths? Sea lettuce? {COVID}-19
  rumors spread like a virus in Japan}.
\newblock
  \bibinfo{howpublished}{\url{https://www.japantimes.co.jp/news/2020/03/08/national/coronavirus-cure-rumors/}}.
\newblock


\bibitem[\protect\citeauthoryear{Prakash, Beutel, Rosenfeld, and
  Faloutsos}{Prakash et~al\mbox{.}}{2012}]%
        {WWW_winner}
\bibfield{author}{\bibinfo{person}{B~Aditya Prakash}, \bibinfo{person}{Alex
  Beutel}, \bibinfo{person}{Roni Rosenfeld}, {and} \bibinfo{person}{Christos
  Faloutsos}.} \bibinfo{year}{2012}\natexlab{}.
\newblock \showarticletitle{Winner takes all: competing viruses or ideas on
  fair-play networks}. In \bibinfo{booktitle}{\emph{Proc. Int. Conf. World Wide
  Web}}. ACM, \bibinfo{pages}{1037--1046}.
\newblock


\bibitem[\protect\citeauthoryear{Rodriguez, Gummadi, and Schoelkopf}{Rodriguez
  et~al\mbox{.}}{2014}]%
        {quantifying_information_overload}
\bibfield{author}{\bibinfo{person}{Manuel~Gomez Rodriguez},
  \bibinfo{person}{Krishna Gummadi}, {and} \bibinfo{person}{Bernhard
  Schoelkopf}.} \bibinfo{year}{2014}\natexlab{}.
\newblock \showarticletitle{Quantifying information overload in social media
  and its impact on social contagions}. In \bibinfo{booktitle}{\emph{Int. Conf.
  Weblogs and Social Media, ICWSM}}. AAAI, \bibinfo{pages}{170--179}.
\newblock


\bibitem[\protect\citeauthoryear{Tang and Liu}{Tang and Liu}{2009}]%
        {BlogCatalog}
\bibfield{author}{\bibinfo{person}{Lei Tang} {and} \bibinfo{person}{Huan Liu}.}
  \bibinfo{year}{2009}\natexlab{}.
\newblock \showarticletitle{Relational learning via latent social dimensions}.
  In \bibinfo{booktitle}{\emph{Proc. Int. Conf. Knowl. Discov. Data Mining,
  KDD.}} \bibinfo{pages}{817--826}.
\newblock


\bibitem[\protect\citeauthoryear{Tricomi}{Tricomi}{1950}]%
        {approx_incomplete_gamma_func}
\bibfield{author}{\bibinfo{person}{FG Tricomi}.}
  \bibinfo{year}{1950}\natexlab{}.
\newblock \showarticletitle{Asymptotische eigenschaften der unvollst{\"a}ndigen
  gammafunktion}.
\newblock \bibinfo{journal}{\emph{Mathematische Zeitschrift}}
  \bibinfo{volume}{53}, \bibinfo{number}{2} (\bibinfo{year}{1950}),
  \bibinfo{pages}{136--148}.
\newblock


\bibitem[\protect\citeauthoryear{Wang, Cui, and Zhu}{Wang
  et~al\mbox{.}}{2016}]%
        {Embedding_link_prediction_2}
\bibfield{author}{\bibinfo{person}{Daixin Wang}, \bibinfo{person}{Peng Cui},
  {and} \bibinfo{person}{Wenwu Zhu}.} \bibinfo{year}{2016}\natexlab{}.
\newblock \showarticletitle{Structural deep network embedding}. In
  \bibinfo{booktitle}{\emph{Proc. Int. Conf. Knowl. Discov. Data Mining, KDD.}}
  \bibinfo{pages}{1225--1234}.
\newblock


\bibitem[\protect\citeauthoryear{Wang and Wang}{Wang and Wang}{2016}]%
        {to_live_or_to_die}
\bibfield{author}{\bibinfo{person}{Jie Wang} {and} \bibinfo{person}{Wenye
  Wang}.} \bibinfo{year}{2016}\natexlab{}.
\newblock \showarticletitle{To live or to die: Encountering conflict
  information dissemination over simple networks}. In
  \bibinfo{booktitle}{\emph{Proc. Conf. Comput. Commun., INFOCOM}}. IEEE,
  \bibinfo{pages}{1--9}.
\newblock


\bibitem[\protect\citeauthoryear{Wei, Valler, Prakash, Neamtiu, Faloutsos, and
  Faloutsos}{Wei et~al\mbox{.}}{2013}]%
        {JSAC_competing_memes}
\bibfield{author}{\bibinfo{person}{Xuetao Wei}, \bibinfo{person}{Nicholas~C
  Valler}, \bibinfo{person}{B~Aditya Prakash}, \bibinfo{person}{Iulian
  Neamtiu}, \bibinfo{person}{Michalis Faloutsos}, {and}
  \bibinfo{person}{Christos Faloutsos}.} \bibinfo{year}{2013}\natexlab{}.
\newblock \showarticletitle{Competing memes propagation on networks: A network
  science perspective}.
\newblock \bibinfo{journal}{\emph{{IEEE} J. Sel. Areas Commun.}}
  \bibinfo{volume}{31}, \bibinfo{number}{6} (\bibinfo{year}{2013}),
  \bibinfo{pages}{1049--1060}.
\newblock


\bibitem[\protect\citeauthoryear{Weng, Flammini, Vespignani, and Menczer}{Weng
  et~al\mbox{.}}{2012}]%
        {scientific_report_limited_attention}
\bibfield{author}{\bibinfo{person}{Lilian Weng}, \bibinfo{person}{Alessandro
  Flammini}, \bibinfo{person}{Alessandro Vespignani}, {and}
  \bibinfo{person}{Fillipo Menczer}.} \bibinfo{year}{2012}\natexlab{}.
\newblock \showarticletitle{Competition among memes in a world with limited
  attention}.
\newblock \bibinfo{journal}{\emph{Scientific Rep.}}  \bibinfo{volume}{2}
  (\bibinfo{year}{2012}), \bibinfo{pages}{335}.
\newblock


\bibitem[\protect\citeauthoryear{{World Health Organization}}{{World Health
  Organization}}{2020a}]%
        {WHO_mythbusters}
\bibfield{author}{\bibinfo{person}{{World Health Organization}}.}
  \bibinfo{year}{(accessed Dec. 9, 2020)}\natexlab{a}.
\newblock \bibinfo{title}{Coronavirus disease ({COVID}-19) advice for the
  public: Mythbusters}.
\newblock
  \bibinfo{howpublished}{\url{https://www.who.int/emergencies/diseases/novel-coronavirus-2019/advice-for-public/myth-busters}}.
\newblock


\bibitem[\protect\citeauthoryear{{World Health Organization}}{{World Health
  Organization}}{2020b}]%
        {COVID19_WHO}
\bibfield{author}{\bibinfo{person}{{World Health Organization}}.}
  \bibinfo{year}{(accessed Dec. 9, 2020)}\natexlab{b}.
\newblock \bibinfo{title}{{WHO} Coronavirus Disease ({COVID}-19) Dashboard}.
\newblock \bibinfo{howpublished}{\url{https://covid19.who.int/table}}.
\newblock


\bibitem[\protect\citeauthoryear{Yang, Yang, and Tang}{Yang
  et~al\mbox{.}}{2017}]%
        {bi_virus_generic_rate}
\bibfield{author}{\bibinfo{person}{Lu-Xing Yang}, \bibinfo{person}{Xiaofan
  Yang}, {and} \bibinfo{person}{Yuan~Yan Tang}.}
  \bibinfo{year}{2017}\natexlab{}.
\newblock \showarticletitle{A bi-virus competing spreading model with generic
  infection rates}.
\newblock \bibinfo{journal}{\emph{{IEEE} Trans. Netw. Sci. Eng.}}
  \bibinfo{volume}{5}, \bibinfo{number}{1} (\bibinfo{year}{2017}),
  \bibinfo{pages}{2--13}.
\newblock


\end{thebibliography}
\normalsize

\end{document}